\documentclass[aps,prd,twocolumn]{revtex4-2}
\usepackage[dvipsnames]{xcolor}
\usepackage[utf8]{inputenc}
\usepackage{amsmath}
\usepackage{stmaryrd}
\usepackage{color}
\usepackage{mathrsfs}
\usepackage{yfonts}
\usepackage{rotating}
\usepackage{graphicx}
\usepackage{amsfonts}
\usepackage{mathrsfs}
\usepackage{amsmath}
\usepackage[american]{babel}
\usepackage{wasysym}
\usepackage{slashed}
\usepackage{pgfplots}
\usepackage{braket}
\usepackage{graphicx}
\usepackage[caption=false]{subfig}
\usepackage{dsfont}
\usepackage{amsthm}
\usepackage{youngtab}
\usepackage{gensymb}
\usepackage{tabulary}

\usepackage{natbib}

\usepackage{mathtools}

\newcommand{\g}[0]{\gamma}
\newcommand{\al}[0]{\alpha}
\newcommand{\be}[0]{\beta}
\newcommand{\de}[0]{\delta}

\newcommand{\dd}[0]{\partial}
\newcommand{\la}[0]{\lambda}
\newcommand{\om}[0]{\omega}
\newcommand{\si}[0]{\sigma}

\newcommand{\bs}[1]{\textbf{#1}}
\newcommand{\bss}[1]{\boldsymbol{#1}}
\newcommand{\ea}[1]{\begin{align}#1\end{align}}
\newcommand{\eq}[1]{\begin{equation}#1\end{equation}}

\newcommand{\ma}[1]{\mathcal{#1}}

\begin{document}
\title{Spacetime Grand Unified Theory}

\author{Gon\c{c}alo M. Quinta}
\email{goncalo.quinta@tecnico.ulisboa.pt}
\affiliation{Quantum Physics of Information Group, Instituto de Telecomunica\c{c}\~{o}es, Lisboa, Portugal}

\begin{abstract}
The gauge symmetries and fermionic sector of the Standard Model are derived from the free Dirac Lagrangian in 8-dimensional spacetime. Motivated by second quantization, we embed the 4-dimensional Clifford algebra of the Dirac Lagrangian into that of 8-dimensional spacetime, showing this embedding carries a redundancy described by a complexified Pati-Salam group acting on field multiplets represented as Clifford algebra ideals, externally and internally via left and right multiplication. Invariance of the Dirac Lagrangian selects different real slices of the group: internal transformations follow $SU(4)_C \times SU(2)_L \times SU(2)_R$, external ones $SU(4)_F \times Spin(1,3)$, with $SU(4)_F$ acting on four particle families. Although $SU(4)_C$ and $SU(4)_F$ can be broken spontaneously and independently, we prove that fixing a triality in $\ma{C}\ell_{8,0}$ simultaneously gives $SU(4)_C \to SU(3)_C \times U(1)_{B-L}$ and $SU(4)_F \to SU(3)_F \times U(1)_{F}$, originating a leptonic sector and a non-mixing fourth family whose stable neutrino is a potential dark matter candidate. Breaking $U(1)_{F}$ yields a hierarchy governed by a generalized Koide formula, with mass scales showing a modular nature. Most importantly, every quantity entering the Standard Model acquires a direct algebraic interpretation in higher dimensional spacetime.
\end{abstract}

\maketitle


\section{Introduction}
\label{Sec1}

Among the unexplained features of the Standard Model (SM), the choice of particle representations and the existence of particle families stand out as particularly mysterious. Grand Unified Theories (GUTs) arose to address some of these problems while unifying the gauge groups of fundamental forces into one framework. The most famous GUTs, based on $SU(5)$ \cite{Georgi:1974a} and $SO(10)$ \cite{Georgi:1974b, Fritzsch:1975}, offered exciting solutions but carried the burden of proton decay, heavily constrained by its experimental lifetime lower limit \cite{Ohlsson:2023}. The Pati–Salam theory \cite{Pati:1974}, though not achieving gauge coupling unification, extended the SM by unifying quarks and leptons and naturally including right-handed neutrinos. Supersymmetry (SUSY) emerged as a promising GUT framework \cite{Golfand:1971,Volkov:1972,Volkov:1973,Wess:1974a,Wess:1974b}, proposing a bosonic (fermionic) superpartner for each fermionic (bosonic) particle. This elegant symmetry resolves the hierarchy problem—the unnatural smallness of the Higgs mass relative to the Planck scale—and offers a dark matter candidate in the lightest neutralino (c.f. \cite{Mohapatra:2003} for an overview of GUTs and SUSY). However, the LHC has found no evidence of SUSY particles within the currently accessible energy ranges. Additional assumptions can circumvent these constraints, but at the cost of new free parameters that undermine the naturalness SUSY was meant to provide.

An algebraic alternative to GUTs and SUSY has been simultaneously developed over the decades, though with far less attention. Typically centered on division algebras \cite{Gunaydin:1973a, Gunaydin:1973b, Dixon:1994, Leo:1996, Carrion:2003, Baez:2012, Baez:2011a, Baez:2011b, Huerta:2011, Huerta:2014, Furey:2015, Burdik:2017, Furey:2021, Furey:2022a, Singh:2023, Furey:2022b, Singh:2025, Furey:2016, Stoica:2018, Furey:2018b, Todorov:2023, Lasenby:2023, Gording:2020, Manogue:2010, Boyle:2014, Boyle:2020, Bhatt:2022, Furey:2014, Furey:2025a,Furey:2025b, Todorov:2017, Furey:2018a, Gillard:2019a, Gillard:2019b} or Clifford algebras \cite{Gresnigt:2026,Barducci:1977, Casalbuoni:1979a, Casalbuoni:1979b, Hestenes:1982, Cho:1995, Chisholm:1996, Lewis:1998, Trayling:2001, Trayling:2004, Doran:2003, Pavsic:2017, Stoica:2020, Borstnik:2023, Zenczykowski:2015, Singh:2025b, Gresnigt:2023,Gourlay:2024,Gresnigt:2026a,Gresnigt:2026b} (c.f. references therein for a fuller list), this approach has yielded intriguing insights. For example, $\ma{C}\ell(7)$ and $\ma{C}\ell(8)$ are known to accommodate one and three copies, respectively, of the SM particle representations \cite{Chisholm:1996,Trayling:2004,Pavsic:2017,Gillard:2019b,Furey:2021}. These works also connect to Geometric Algebra (c.f. \cite{Doran:2003} and references therein), which aims to link SM symmetries to transformations on the Clifford algebra of spacetime. On the division algebra front, the possible relation between quaternions and octonions to SM symmetries has likewise been explored for decades \cite{Gunaydin:1973a, Gunaydin:1973b, Dixon:1994, Leo:1996, Carrion:2003, Baez:2012, Baez:2011a, Baez:2011b, Huerta:2011, Huerta:2014, Furey:2015, Burdik:2017, Furey:2021, Furey:2022a, Singh:2023, Furey:2022b, Singh:2025, Furey:2016, Stoica:2018, Furey:2018b, Todorov:2023}, often by choosing objects such as octonion unit elements or volume forms as invariant structures selecting SM symmetries within larger transformations. Despite their merit, these approaches are often limited by the ad hoc nature of the structures involved and lack testable predictions. Further work has explored particle families via possible connections with triality \cite{Silagadze:1995, YuFen:2006, Manogue:2010, Boyle:2014, Furey:2014, Todorov:2017, Furey:2018a, Gillard:2019a, Gillard:2019b, Gording:2020, Boyle:2020, Bhatt:2022, Furey:2025a, Furey:2025b}, though no consensus picture exists yet, partly because triality lacks a principled motivation in this context, undermining testable predictions. Still, some works derive phenomenological consequences for the Yukawa matrices \cite{Singh:2025b, Gresnigt:2026b}. Most of these algebraic approaches, however, overlook the Quantum Field Theory (QFT) aspect (notable exceptions being \cite{Cho:1995,Zenczykowski:2015,Pavsic:2017}).

Here we depart from the historical top-down search for mathematical frameworks containing the SM gauge symmetries, taking a bottom-up approach instead: we show that the SM can be obtained solely from the free Dirac QFT. The algebra of spacetime (the Dirac algebra) becomes unified with the algebra of gauge transformations, revealing a deeper spacetime origin of fundamental forces. Ideals and Lie algebras become more than entities associated to abstract groups, connecting directly to quantum field multiplets which transform as (not just under) Clifford algebras.

The paper is organized as follows. Sec.~\ref{Sec2} uses second quantization of the Dirac equation to motivate generalizing the complexified Dirac algebra $\ma{C}\ell_{1,3}$ to $\ma{C}\ell_{8,0}$, introducing 8-dimensional spacetime. Sec.~\ref{Sec3} shows that 8-dimensional spacetime accommodates four copies of 4-dimensional Dirac spinors, giving a generalized free Dirac Lagrangian in higher dimensions. Sec.~\ref{Sec4} proves that particle multiplets are naturally associated to Clifford algebra ideals and finds the corresponding gauge symmetries and particle representations of the theory. Sec.~\ref{Sec5} deals with the spontaneous symmetry breaking of the internal symmetries described by the Pati-Salam group, focusing on the higher dimensional spacetime interpretation of all related quantities. Sec.~\ref{Sec6} repeats the latter task for external symmetries and develops the phenomenological implications for the Yukawa matrices. In Sec.~\ref{Sec7} we provide a new parameterization of the family masses and show that mass scales have an inherent modular nature. We finish with concluding remarks in Sec.~\ref{Sec8}.

\section{Second quantized operators as Clifford algebra elements}
\label{Sec2}

We begin by considering a free fermionic quantum field $\psi(x)$ with mass $m$, living in a Minkowski spacetime with metric $\eta = \textrm{diag}(+1,-1,-1,-1)$ and parametrized by coordinates $x^{\mu}$. Such a theory is governed by the Lagrangian operator
\eq{\label{FreeLagrangian}
\ma{L}(x) = \psi^{\al \dagger}(x)(i(\g^{0}\g^{\mu})_{\al\be}\dd_{\mu}-m (\g^{0})_{\al\be})\psi^{\be}(x)
}
where $\g^{\mu}$ are the Dirac gamma matrices. We assume the usual Einstein notation whenever summations are not explicit. The gamma matrices obey
\eq{\label{DiracAlgebra}
\{\g^{\mu},\g^{\nu}\} = 2\eta^{\mu\nu} 1_{4\times 4}\,,
}
with $1_{n \times n}$ the n-dimensional identity matrix, generating $\ma{C}\ell_{1,3}$ (complex Clifford algebras $\mathbb{C} \otimes \ma{C}\ell_{p,q}$ are implicit unless otherwise stated). We adopt $\ma{C}\ell_{r,q}$ for a Clifford algebra generated by $r$ elements of norm $+1$ and $q$ of norm $-1$, or $\ma{C}\ell(n)$ for an unspecified $n$-dimensional metric. Since the $\g^{\mu}$ transform as 4-vectors with an inner product defined by Eq.~(\ref{DiracAlgebra}), we will frequently call $\g^{\mu}$ by spacetime basis vectors, as they are indeed a representation of spacetime basis vectors in $\ma{C}\ell_{1,3}$. Physically, the number of generators of $\ma{C}\ell_{1,3}$ can be read as the assumed spacetime dimensionality, an interpretation we adopt throughout. Of special significance for this work are the creation and annihilation operators $a^{s}_{\bs{p}}$ and $b^{s}_{\bs{p}}$ present in $\psi(x)$, which obey the anticommutation relations $\{a^{s}_{\bs{p}}, a^{r \dagger}_{\bs{k}}\} = \{b^{s}_{\bs{p}}, b^{r \dagger}_{\bs{k}}\} = (2\pi)^3 \de^{sr}\de^{(3)}(\bs{p}-\bs{k})$ and $\{a^{s}_{\bs{p}}, a^{r}_{\bs{k}}\} = \{a^{s \dagger}_{\bs{p}}, a^{r \dagger}_{\bs{k}}\} = \{b^{s}_{\bs{p}}, b^{r}_{\bs{k}}\} = \{b^{s \dagger}_{\bs{p}}, b^{r \dagger}_{\bs{k}}\} = 0$. Focusing on the spin indexes (i.e. on a single momentum mode), the latter relations resemble a Witt basis of $\ma{C}\ell(8)$
\ea{\label{WittCl8}
\{\om^{\al},\om^{\be \dagger}\} = \de^{\al\be}\,, \quad \{\om^{\al},\om^{\be}\} = \{\om^{\al \dagger},\om^{\be \dagger}\} = 0\,,
}
with $\al,\be = 1,\ldots, 4$, built from a basis $\{e^k\}_{k=1,\ldots,8}$ (i.e. $\{e^i,e^k\} = 2\de^{ik}$, akin to Eq.~(\ref{DiracAlgebra})) via \cite{Lounesto:2001}
\eq{\label{explicitWitt}
\om^{\al} = \frac{i e^{\al} - e^{9-\al}}{2}\,, \quad \al=1,\cdots,4\,,
}
where we explicitly consider the complexification of $\ma{C}\ell_{8,0}$ and use the Witt basis metric in (\ref{WittCl8}) to convert between upper and lower indexes (equivalent here). Thus, focusing on a single momentum mode, an 8-dimensional Clifford algebra is the smallest one fully describing a second quantized Dirac spinor, since it is the smallest algebra with exactly four creation/annihilation operators as Witt basis vectors. A second quantized Dirac theory then follows by building fermionic creation/annihilation operators as infinite tensor products of $\ma{C}\ell_{8,0}$ algebras (c.f. Appendix \ref{App1}), with $a^{s}_{\bs{p}}, b^{r}_{\bs{q}} \in \ma{C}\ell_{8,0}\otimes \lim_{N \to \infty} (\ma{C}\ell_{8,0})^{\otimes 3N}$ as infinite dimensional matrices. Therefore, just as relativistic invariance of quantum dynamics motivates the introduction of $\ma{C}\ell_{1,3}$ and the Dirac equation, quantizing its four independent solutions motivates the generalization to $\ma{C}\ell_{8,0}$. While the full Dirac QFT does not uniquely single out $\ma{C}\ell_{8,0}$, the single-momentum-mode argument will be considered compelling enough to carry this generalization, which can be physically interpreted as considering the algebra of 8-dimensional spacetime instead of a 4-dimensional one. Higher dimensional algebras are not ruled out, but we shall see that $\ma{C}\ell_{8,0}$ suffices to incorporate the SM.

\section{Dirac spinors in eight dimensions}
\label{Sec3}

Having motivated $\ma{C}\ell_{8,0}$, we may ask whether it plays a further role in the theory of Eq.~(\ref{FreeLagrangian}), equivalently written as $\ma{L}(x) = \psi^{\al \dagger}(x) X_{\al\be} \psi^{\be}(x)$ with $X_{\al\be}$ the components of a $4\times 4$ matrix $X \in \ma{C}\ell_{1,3}$. It is natural to seek an embedding of the Dirac algebra $X\in\ma{C}\ell_{8,0}$ whose lowest matrix representation is 16-dimensional. The task is then to find four 16-dimensional matrices $\Gamma^{\mu}_{(+)}$ and a 16-dimensional diagonal matrix $P_{(+)}$ with only four non-zero entries, each equal to 1, such that
\eq{\label{DiracAlgebra8D}
\{\Gamma^{\mu}_{(+)},\Gamma^{\nu}_{(+)}\} = 2 \eta^{\mu\nu} P_{(+)}\,,
}
where $(+)$ is a label kept for later convenience. For a natural representation, consider the four annihilation operators of a single momentum mode $a^{1}_{\bs{p}}=\om^{1},a^{2}_{\bs{p}}=\om^{2}, b^{1}_{\bs{p}} = \om^{3}, b^{2}_{\bs{p}} = \om^{4} \in \ma{C}\ell_{8,0}$ in a lattice, focusing on the Clifford algebra tied to the spinorial quantization degrees of freedom (c.f. Eqs.~(\ref{a1})-(\ref{b2}) in the Appendix). Defining
\eq{
\left(V_A\right)_{ij} = \de_{i,A}\de_{j,14}\,, \quad i,j=1,\ldots,16
}
we find $a^{s}_{\bs{p}} V^{\dagger}_A = b^{s}_{\bs{p}} V^{\dagger}_A = 0\,, a^{s \dagger}_{\bs{p}} V^{\dagger}_A \neq 0\,, b^{s \dagger}_{\bs{p}} V^{\dagger}_A \neq 0$ for the representation (\ref{eAbasis}) used here, i.e. $V^{\dagger}_{A}$ is a vacuum for any $A$, with the QFT vacuum built as tensor products of it (c.f. Eq.~(\ref{QFTvac}) in Appendix \ref{App1}). Since $XV^{\dagger}_{A}\, (\forall_{X\in\ma{C}\ell_{8,0}})$ is a matrix with a non-zero column $A$ and 0 elsewhere, $V^{\dagger}_{A}$ are minimal left ideals generating independent spaces under left multiplication. Dirac spinors thus generalize to $\ma{C}\ell_{8,0}$ as matrices of the form $XV^{\dagger}_{A}$. For example, $a^{1 \dagger}_{\bs{p}=0}V^{\dagger}_{A}$ can consistently be read as creating a spinor $u^{1}(0)$ in 8 dimensions, since it is non-zero only in column $A$, with a single entry equal to 1. This motivates identifying Dirac rest spinors in 8-dimensional spacetime as
\ea{
u^{1}_{8D}(\bs{p}=0) & = a^{1 \dagger}_{\bs{p}=0}V^{\dagger}_{A} \equiv \om^{1 \dagger}V^{\dagger}_{A}\,, \label{u18D} \\
u^{2}_{8D}(\bs{p}=0) & = a^{2 \dagger}_{\bs{p}=0}V^{\dagger}_{A} \equiv \om^{2 \dagger}V^{\dagger}_{A}\,, \label{u28D} \\
v^{1}_{8D}(\bs{p}=0) & = b^{1 \dagger}_{\bs{p}=0}V^{\dagger}_{A} \equiv \om^{3 \dagger}V^{\dagger}_{A}\,, \label{v18D} \\
v^{2}_{8D}(\bs{p}=0) & = b^{2 \dagger}_{\bs{p}=0}V^{\dagger}_{A} \equiv \om^{4 \dagger}V^{\dagger}_{A}\,. \label{v28D}
}
While the connection between the Witt basis of $\ma{C}\ell_{8,0}$ and creation-annihilation operators has been noted previously (c.f. \cite{Borstnik:2023} and references therein), unifying the algebra of a single mode's creation operators with that of the spinors they create has never been realised before. This could not be done in $\ma{C}\ell_{1,3}$, whose Witt basis has only two independent vectors, so $\ma{C}\ell_{8,0}$ is the simplest algebra unifying single-mode creation operators and Dirac spinors, reinforcing its importance. A representation for $\Gamma_{(+)}^{\mu}$ then follows by first recalling the completeness relation in the rest frame \cite{Bjorken:1964} $1_{4 \times 4} = \sum^{2}_{s=1} u^{s}(0) u^{s \dagger}(0) + v^s(0) v^{s \dagger}(0)$, whose 8-dimensional version
\ea{
\sum^{2}_{s=1} \bigg(u_{8D}^{s}(0) u_{8D}^{s \dagger} + v_{8D}^s(0) v_{8D}^{s \dagger}(0)\bigg) & \equiv P_{(+)} \nonumber \\
& = \sum^{4}_{\al=1} \om^{\al \dagger} \ma{V} \om^{\al} \label{completeId}
}
can be obtained using Eqs.~(\ref{u18D})-(\ref{v28D}), where we defined
\eq{
\ma{V} = V_{14} V^{\dagger}_{14}
}
and used $V^{\dagger}_A V_A  = \ma{V}$. For the basis $\{e^k\}$ of $\ma{C}\ell_{8,0}$ in Eq.~(\ref{eAbasis}), we have $P_{(+)} = \textrm{diag}(0,0,0,0,0,0,0,0,1,1,1,1,0,0,0,0)$, which is independent from the rest spinors in Eq.~(\ref{u18D})-(\ref{v28D})), i.e. from the basis chosen to express $\g^{\mu}$.  Eq.~(\ref{completeId}) shows that $P_{(+)}$ is an embedding of the 4-dimensional identity matrix in $\ma{C}\ell_{8,0}$, so any representation of $\Gamma^{\mu}_{(+)}$ must satisfy ${[\Gamma^{\mu}_{(+)},P_{(+)}] = 0}$ which is trivially verified for $\Gamma^{\mu}_{(+)} = P_{(+)}\Gamma^{\mu}_{(+)}P_{(+)}$. Since $\Gamma^{\mu}_{(+)}$ generates all elements of the Dirac algebra, any element $X$ of such algebra must satisfy ${X = P_{(+)} X P_{(+)}}$. On the other hand, Eq.~(\ref{completeId}) implies that $\{\om^{\al \dagger} \ma{V} \om^{\be}\}$ is a basis for embedding 4-dimensional matrices in 16-dimensional ones. Consequently, if $X_{\al\be}$ are the components of a Dirac algebra element in 4-dimensions, then
\eq{\label{8Dembbed}
X = \sum^{4}_{\al,\be=1} X_{\al\be} \, \om^{\al \dagger} \ma{V} \om^{\be}
}
is a natural embedding of such an element in $\ma{C}\ell_{8,0}$. Finally, Eq.~(\ref{completeId}) leads to $\om^{\al \dagger} \ma{V} = P_{(+)} \om^{\al \dagger}$ which can be used in Eq.~(\ref{8Dembbed}) to obtain the alternative embedding
\eq{\label{X8D}
X = P_{(+)} X_{(\om)} P_{(+)}\,, \quad X_{(\om)} = \sum^{4}_{\al,\be=1} X_{\al\be} \, \om^{\al \dagger} \om^{\be}\,.
}
Eq.~(\ref{X8D}) embeds matrix components of the Lagrangian as elements of $\ma{C}\ell_{8,0}$ by inverting Eq.~(\ref{8Dembbed}), using Eq.~(\ref{WittCl8}), $\Gamma^{\mu}_{(+)} = P_{(+)}\Gamma^{\mu}_{(+)}P_{(+)}$, $\om^{\al \dagger} \ma{V} = P_{(+)} \om^{\al \dagger}$ and $\ma{V}^2 = \ma{V}$, leading to
\eq{\label{XalbeV1}
X_{\al\be} = \textrm{Tr}[\om_{\al} X_{(\om)} \om^{\dagger}_{\be}\ma{V}]\,,
}
where $\textrm{Tr}$ denotes the standard matrix trace. A key consequence of Eq.~(\ref{XalbeV1}) is the automatic inheritance by the Lagrangian of properties characteristic to $\ma{C}\ell_{8,0}$. One such example is the exotic phenomenon of triality, which asserts that the two spinorial representations $S_{\pm}$ of Spin(8) and the vectorial one $V$ are interrelated due to the coincidental matching of dimensionalities. Together with the three fundamental and adjoint representations of Spin(8), triality implies that Eq.~(\ref{XalbeV1}) can be expressed in four independent ways interconnected in a particularly symmetric fashion, each with a multiplicity of sixteen due to $\ma{V} = V^{\dagger}_A V_A, \forall A$. For the matrices $(\g^0 \g^{\mu})_{\al\be}$ and $(\g^0)_{\al\be}$ appearing in (\ref{FreeLagrangian}), one finds in particular (c.f. Appendix \ref{App2})
\ea{
(\g^0 \g^{\mu})_{\al\be} & = \textrm{Tr}\left[V_A \xi^{\dagger}_{\al,g} \Gamma^{0}\Gamma^{\mu} \xi_{\be,g} V^{\dagger}_A\right]\,, \label{g0gmuTrFinal} \\
(\g^0)_{\al\be} & = \textrm{Tr}\left[V_A \xi^{\dagger}_{\al,g} \Gamma^{0} \xi_{\be,g} V^{\dagger}_A\right]\,, \label{g0TrFinal}
}
where $\Gamma^{\mu} = 1_{4\times 4} \otimes \gamma^{\mu}$ generate a Dirac algebra, the objects
\ea{
\xi_{\al,1} & = \om^{\dagger}_{\al} \Omega_{0}\,, \label{xi1} \\
\xi_{\al,2} & = \om^{\dagger}_{\al} \Omega_{+}\,, \label{xi2} \\
\xi_{\al,3} & = \om^{\dagger}_{\al} \Omega_{-}\,, \label{xi3} \\
\xi_{\al,4} & = \kappa_{\al} \label{xi4}
}
are four ideals, denoted as \textit{Dirac ideals}, involving three linear independent ideals $\Omega_{a}$ ($a\in\{0,+,-\}$) satisfying the properties ${\om_{\al} \Omega^{\dagger}_{a} = 0}$, ${\om^{\dagger}_{\al} \Omega^{\dagger}_{a} \neq 0}$ and ${\Omega^{\dagger}_{a}\Omega_{b} = 2\de_{ab} \ma{V}}$, and
\eq{
\kappa_{1} = V_{13}\,, \, \kappa_{2} = V_{14}\,, \, \kappa_{3} = V_{15}\,, \, \kappa_{4} = V_{16}
}
are fixed as functions of $\Omega_{a}$ and $\om_{\al}$. For future convenience, we shall denote $\Omega_{a}$ as \textit{family ideals}, explicitly given by
\ea{
\Omega_{+} & = \left(\frac{1+w}{8}\right)\left(\frac{1+I_e}{2}\right)(2 n\om^{\dagger}_{1})\,, \label{OmPlus} \\
\Omega_{-} & = \left(\frac{1+w}{8}\right)\left(\frac{1-I_e}{2}\right)(2 \om^{\dagger}_{1})\,, \label{OmMinusVac} \\
\Omega_{0} & = (-\om_1\om_2+\om_3\om_4)\om^{\dagger}_{1}\om^{\dagger}_{2}\om^{\dagger}_{3}\om^{\dagger}_{4}\,, \label{OmVec}
}
where $I_e = e_1\cdots e_8$ is the 8-dimensional pseudoscalar, $n=e_8$ and $w=e_1 e_2 e_3 e_4 + e_1 e_4 e_6 e_7 - e_1 e_3 e_5 e_7 + e_1 e_2 e_5 e_6 - e_3 e_4 e_5 e_6 - e_2 e_4 e_5 e_7 - e_2 e_3 e_6 e_7$. The object $w$ is commonly known as a calibration and $n$ its neutral axis. The specific choices Eq.~(\ref{OmPlus})-(\ref{OmVec}) are not unique (c.f. Appendix \ref{App2}). The four ideals $\xi_{\al,g}$ are associated to each of the four Spin(8) representations described by the respective $D_4$ Dynkin diagram, as depicted in Fig.~1.
\begin{figure}[h!]
  \centering
  \includegraphics[width=0.4\textwidth]{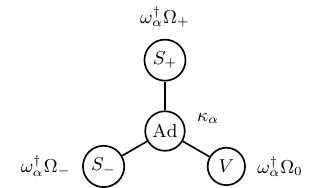}
  \caption{The four possible Dirac ideal representations are associated to each of the representations in the Spin(8) Dynkin diagram, displaying a $D_4$ symmetry once a calibration and neutral axis are fixed.}
  \label{fig:1}
\end{figure}
The first three $\xi_{\al,a}$ $(a=1,2,3)$ are associated to the vector and two spinorial representations while $\xi_{\al,4}$ is related to the adjoint representation. Most importantly, $\xi_{\al,4} = \kappa_{\al}$ cannot be expressed as $\om^{\dagger}_{\al}\Omega_{4}$ for any $\Omega_{4} \in \ma{C}\ell_{8,0}$, setting it apart from the symmetric form of the remaining representations. Using the representation in (\ref{eAbasis}), a linear combination of $\xi_{\al,g}$ is a matrix with a single non-zero column
\ea{\label{generalSpinor}
\Psi & = \sum_{\al,g} \psi_{\al,g} \, \xi_{\al,g} = \Psi \ma{V} \nonumber \\
& = (\psi_{(1)}, \psi_{(2)}, \psi_{(3)}, \psi_{(4)})^T \, \bs{e}_{14} \,,
}
where
\eq{
\psi_{(g)} \equiv (\psi_{1,g}, \psi_{2,g}, \psi_{3,g}, \psi_{4,g})\,,  \quad \psi_{\al,g} \in \mathbb{C}
}
and where $\bs{e}_k$ are the standard euclidean row basis vectors of $\mathbb{R}^{16}$. In other words, $\Psi$ is a direct sum of four independent 4-dimensional ideals, each one associated to a Spin(8) representation once a calibration and neutral axis are fixed, a process which will be equivalently referred to as ``fixing a triality''. Multiplying $\Psi$ from the right by $V^{\dagger}_A$ will shift the non-zero column towards column $A$. Consequently, the combination
\eq{
W_{\Psi} = \sum^{4}_{\al=1} \sum^{4}_{g=1} \sum^{16}_{A=1} \psi_{\al,g,A} \, \xi_{\al,g} V^{\dagger}_A \,, \quad \psi_{\al,g,A} \in \mathbb{C}
}
represents a general element of $\ma{C}\ell_{8,0}$. The projectors $P_{(g)} = \sum_{\al} \xi_{\al,g} \xi^{\dagger}_{\al,g}$ correspond to the identity elements in four mutually orthogonal 4-dimensional subspaces of $C\ell_{8,0}$, where Lorenztian spacetimes can be identified via the construction of Dirac algebras $C\ell_{1,3}$, each one spanned by $\Gamma^{\mu}_{(g)} = P_{(g)} \Gamma^{\mu} P_{(g)}$ (hence the notation of Eq.~(\ref{DiracAlgebra8D})).

\section{Gauge transformations from ideal rotations}
\label{Sec4}

We now identify the redundancies of the previous construction, which will be naturally identified as the gauge symmetries of the theory. Each pair $(g,A)$ in Eqs.~(\ref{g0gmuTrFinal})-(\ref{g0TrFinal}) gives the same Lagrangian matrix elements, exposing redundant indexes absent from the algebra of 4-dimensional spacetime. Physically no pair is preferred, so each should be associated to an independent quantum field $\psi_{\al,g,A}(x)$ obeying the usual Fermi anticommutation rules. For a specific particle type $(g,A)$, this means inserting Eqs.~(\ref{g0gmuTrFinal})-(\ref{g0TrFinal}) into Eq.~(\ref{FreeLagrangian}) and summing the generalized terms over all independent fields. For the kinetic term, this gives
\ea{
\ma{L}^{\textrm{Kin.}}(x) & = \sum_{g,A} \sum_{\al,\be}\psi^{\dagger}_{\al,g,A}(x)(i\g^0\g^{\mu})_{\al\be}\dd_{\mu}\psi_{\be,g,A}(x) \nonumber \\
& = \textrm{Tr}\bigg[W_{\Psi}(x)^{\dagger} (i\Gamma^{0}\Gamma^{\mu}) \dd_{\mu} W_{\Psi}(x)\bigg]\,, \label{LkinaA}
}
where we define
\eq{\label{MultipletW}
W_{\Psi}(x) = \sum^{4}_{\al=1} \sum^{4}_{g=1} \sum^{16}_{A=1} \psi_{\al,g,A}(x) \xi_{\al,g} V^{\dagger}_A\,.
}
Eq.~(\ref{MultipletW}) makes it clear that field multiplets are naturally identified with $\xi_{\al,g} V^{\dagger}_A$. In addition, there is a bijection $A\equiv A(\al,g)$ such that $V_{A(\al,g)} = \xi_{\al,g}$, so each $V_A$ equals one particular ideal $\xi_{\al,g}$, and so the gauge symmetries of the theory can be identified by focusing solely on the ideals $\xi_{\al,g}$ and their rotations.

Before examining the gauge transformations explicitly, note that the indexes in $\xi_{\al,g} V^{\dagger}_A$ respond differently to multiplication from each side: a general matrix from the left affects only $\al$ and $g$, while one from the right changes only $A$. External transformations, which affect spinor indexes, are thus left multiplications in (\ref{MultipletW}), while internal transformations, blind to spinor indexes, are right multiplications. This identifies $A$ as flavour indexes and $g$ as family indexes, predicting four particle families. External and internal symmetries are thus unified as objects of the same algebra while avoiding the Coleman-Mandula theorem \cite{Coleman:1967} through action on disconnected spaces. Since left multiplications also affect family indexes generally, family transformations may be linked to external transformations in the extra spacetime dimensions.

Regarding the explicit form of the gauge transformations, since all field multiplets inherit the rotation properties of $\xi_{\al,g}$, we need only to fully characterize the latter. In particular, gauge transformations are associated to mathematically redundant degrees of freedom, which are naturally present in this theory as changes of basis parameterized by individual rotations of the indexes $\al$ and $g$. The former are spinor indexes whose rotations generally live in the group $Spin(1,3)_{\mathbb{C}}$, while the latter are indexes of complex components whose rotations live in $SL(4,\mathbb{C})$. In light of the tensor product structure
\eq{
\sum_{\al,g} c_{\al} d_{g} \xi_{\al,g} =
\begin{pmatrix}
0 & 0 & 0 & d_1 \\
0 & 0 & 0 & d_2 \\
0 & 0 & 0 & d_3 \\
0 & 0 & 0 & d_4
\end{pmatrix} 
\otimes
\begin{pmatrix}
0 & c_1 & 0 & 0 \\
0 & c_2 & 0 & 0 \\
0 & c_3 & 0 & 0 \\
0 & c_4 & 0 & 0
\end{pmatrix}
}
we find that, on purely algebraic grounds, rotations of $\xi_{\al,g}$ are parameterized by
\eq{\label{genXirot}
\xi'_{\al,g} = e^{X_{\om}+X}\xi_{\al,g}\,,
}
with the minus sign introduced for convenience, where the generators
\ea{
X_\om & = a_{\mu\nu} \Sigma^{\mu\nu}\,, \quad a_{\mu\nu} \in \mathbb{C} \\
\Sigma^{\mu\nu} & = [\Gamma^{\mu},\Gamma^{\nu}] = 1_{4\times 4}\otimes([\gamma^{\mu},\gamma^{\nu}]) \label{LorentzGens}
}
span $Spin(1,3)_{\mathbb{C}}$ and
\eq{
X = l_{k} T^{k} \otimes 1_{4 \times 4}\,, \,\,\, l_i \in \mathbb{C}\,, \\
 \label{SU4gens}
}
span $SL(4,\mathbb{C})$ ($T^i$ are the generalized Gell-Mann matrices of $SU(4)$). In particular, due to the isomorphism
\eq{
Spin(1,3)_{\mathbb{C}} \simeq Spin(4)_{\mathbb{C}} \simeq SL(2,\mathbb{C})\times SL(2,\mathbb{C})
}
we find that
\eq{\label{genXigroup}
e^{X_{\om}+X} \in SL(4,\mathbb{C})\times SL(2,\mathbb{C})\times SL(2,\mathbb{C}) \equiv G
}
or, in other words, the group of rotations of $\xi_{\al,g}$ associated to changes of basis is the complexified Pati-Salam group \cite{Pati:1974}. We now proceed to fix a triality in $\ma{C}\ell_{8,0}$ by choosing a calibration and neutral axis. Using the choices from the previous section, we are led to Eqs.~(\ref{xi1})-(\ref{xi4}), which clearly only allow rotations between the indexes $g=1,2,3$. Any generator of $X$ leading to a rotation mixing $g=4$ with the remaining indexes will correspond to a different choice of pair $(w,n)$ and is thus prohibited if the chosen triality of Spin(8) is to be conserved. This way, $X$ is broken down to
\ea{ 
X_{\Omega} & = c_{a} \Lambda^{a} + c_0 Q_{B-L}\,, \quad c_{a}, c_0 \in \mathbb{C} \label{XOmega} \\
\Lambda^{a} & =
\begin{pmatrix}
  & & & 0 \\
  & \lambda^a & & 0 \\
  & & & 0 \\
  0 & 0 & 0 & 0
\end{pmatrix} \otimes 1_{4\times 4}\,, \label{SU3gens} \\
Q_{B-L} & =
\begin{pmatrix}
  \frac{1}{3} & 0 & 0 & 0 \\
  0 & \frac{1}{3} & 0 & 0 \\
  0 & 0 & \frac{1}{3} & 0 \\
  0 & 0 & 0 & -1
\end{pmatrix} \otimes 1_{4\times 4}\,, \label{Qgens}
}
where $\lambda^{a}$ are the Gell-Mann matrices, which is mathematically equivalent to taking different linear combinations of the family ideals $\Omega_a$ in $\xi_{\al,a}$. As we shall see in the next section, the breaking of $SU(4)$ induced by fixing a triality can also be obtained dynamically from spontaneous breaking. The point is to show that an algebraic condition could also be an origin for this breaking without invoking extra scalar fields.

From a dynamical point of view, there are further restrictions to implement. Focusing on general external transformations of field multiplets involving Eq.~(\ref{genXirot})
\eq{\label{InternalT}
W'_{\Psi}(x) = e^{X_{\om}+X}W_{\Psi}(x')\,,
}
we find that the Lagrangian (\ref{LkinaA}) is only invariant under certain conditions. Firstly, since $[X,\Gamma^{\mu}]=0$, one must have $X^{\dagger} = - X$. For the $Spin(1,3)_{\mathbb{C}}$ rotation coming from the generator $X_{\om}$, a cancellation in the Lagrangian is only possible by restricting the corresponding generators to pure real coefficients, followed by a Lorentz transformation in the spacetime coordinates. In other words, the rotations in (\ref{InternalT}) are restricted to
\eq{
W'_{\Psi}(x) = e^{S_{\om}+i H}W_{\Psi}(x')
}
where
\eq{
S_{\om} = d_{\mu\nu} \Sigma^{\mu\nu}\,, \quad d_{\mu\nu} \in \mathbb{R} \\
}
are the spinor generators of $Spin(1,3)$, $x'= \Lambda x$ is the Lorentz transformation associated to $S_{\om}$ and
\eq{
H = t_{k} T^{k} \otimes 1_{4 \times 4}\,, \quad t_{k} \in \mathbb{R}
}
generates a $SU(4)_F$ symmetry acting on the four family indexes. If a triality is fixed, or a corresponding spontaneous breaking occurs, the generators of $H$ are reduced to (\ref{SU3gens}) and (\ref{Qgens}) generating a $SU(3)_F\times U(1)_F$ symmetry, where the abelian component is generated by $Q_{B-L}$. The most general group of external rotations leaving the Lagrangian invariant once a triality is fixed is thus
\eq{
G_E = Spin(1,3) \times U(3)_F\,,
}
where
\eq{
U(3)_F = SU(3)_F\times U(1)_F\,.
}
The first three families each transform as a $\overline{\bs{3}}$ under $SU(3)_F$ with charges $1/3$ under $U(1)_F$ while the fourth family transforms as a singlet with charge $-1$. Under internal rotations of field multiplets
\eq{
W'_{\Psi}(x) = W_{\Psi}(x)e^{X^{\dagger}_{\om}+X^{\dagger}}\,,
}
the Lagrangian (\ref{LkinaA}) is only invariant if $X^{\dagger}_{\om}=-X_{\om}$ and $X^{\dagger}=-X$. This amounts to taking the unitary piece of the group in Eq.~(\ref{genXigroup}), corresponding to the Pati-Salam group $SU(4)_C\times SU(2)_L \times SU(2)_R$. Fixing a triality then leads to the restricted group
\eq{
G_I = SU(3)_C \times SU(2)_L \times SU(2)_R \times U(1)_{B-L}\,,
}
which could also be obtained via the standard spontaneous breaking of $SU(4)_C$, detailed in the next section. While the generators of $SU(3)_C\times U(1)_{B-L}$ remain the same as (\ref{XOmega}), the generators of $SU(2)_L \times SU(2)_R$ are
\ea{
T^{i}_{L} & = \frac{1}{2} P_{L} \Gamma^{0}\Gamma^{i} P_{L} = 1_{4\times 4} \otimes \left(0_{2\times 2} \oplus \frac{\sigma^i}{2}\right) \,, \label{TL} \\
T^{i}_{R} & = \frac{1}{2} P_{R} \Gamma^{0}\Gamma^{i} P_{R} = 1_{4\times 4} \otimes \left(-\frac{\sigma^i}{2} \oplus 0_{2\times 2}\right) \,, \label{TR}
}
with the chiral projection operators
\ea{\label{PLdef}
P_L = \frac{1+\Gamma^5}{2}\,, \quad P_R = \frac{1-\Gamma^5}{2}
}
and where
\eq{\label{Gamma5}
\Gamma^5 = i \Gamma^0\Gamma^1\Gamma^2\Gamma^3 = 1_{4\times 4} \otimes \gamma^5
}
is the 4-dimensional pseudoscalar. Note the (unphysical) relative sign change between $T^{i}_{L}$ and $T^{i}_{R}$ appearing due to the choice of maintaining the symmetric form of both sets of generators. Since $P_{L,R}$ are also used to select chiral components of spinors via left multiplication, we find that internal and external chiralities are unified. Having defined all generators of the groups $G_I$ and $G_E$, we can identify the particle representations which are summarized in Tables~\ref{Table1} and \ref{Table2}. In addition to the fermionic particle content of the SM, we find right-handed neutrinos and a fourth particle family. The same generators are used for colour and family degrees of freedom, which are physically decoupled by the independent action of the group (\ref{genXirot}) via right and left multiplication, respectively. From an external viewpoint, the $Q_{B-L}$ generator singles out a non-mixing family, while it leads to the colourless leptonic sector from an internal point of view. In addition, the present particle content leads to a full anomaly cancellation.
\begin{table}[h!]
\centering
\begin{tabular}{|c|c|c|c|c|c|}
\hline
\multicolumn{6}{|c|}{Representations of $\xi_{\alpha,g} V^{\dagger}_A$ under right action of $G_{I}$} \\
\hline
Ideal    & $SU(3)_C$    & $SU(2)_L$             & $SU(2)_R$             & $U(1)_{B-L}$ & Flavour    \\
\hline
$V^{\dagger}_1$    & $\bs{3}^r$ & $\bs{1}$              & $\bs{2}^{\downarrow}$ & $1/3$        & $d^r_R$    \\
$V^{\dagger}_2$    & $\bs{3}^r$ & $\bs{1}$              & $\bs{2}^{\uparrow}$   & $1/3$        & $u^r_R$    \\
$V^{\dagger}_3$    & $\bs{3}^r$ & $\bs{2}^{\uparrow}$   & $\bs{1}$              & $1/3$        & $u^r_L$    \\
$V^{\dagger}_4$    & $\bs{3}^r$ & $\bs{2}^{\downarrow}$ & $\bs{1}$              & $1/3$        & $d^r_L$    \\
$V^{\dagger}_5$    & $\bs{3}^g$ & $\bs{1}$              & $\bs{2}^{\downarrow}$ & $1/3$        & $d^g_R$    \\
$V^{\dagger}_6$    & $\bs{3}^g$ & $\bs{1}$              & $\bs{2}^{\uparrow}$   & $1/3$        & $u^g_R$    \\
$V^{\dagger}_7$    & $\bs{3}^g$ & $\bs{2}^{\uparrow}$   & $\bs{1}$              & $1/3$        & $u^g_L$    \\
$V^{\dagger}_8$    & $\bs{3}^g$ & $\bs{2}^{\downarrow}$ & $\bs{1}$              & $1/3$        & $d^g_L$    \\
$V^{\dagger}_9$    & $\bs{3}^b$ & $\bs{1}$              & $\bs{2}^{\downarrow}$ & $1/3$        & $d^b_R$    \\
$V^{\dagger}_{10}$ & $\bs{3}^b$ & $\bs{1}$              & $\bs{2}^{\uparrow}$   & $1/3$        & $u^b_R$    \\
$V^{\dagger}_{11}$ & $\bs{3}^b$ & $\bs{2}^{\uparrow}$   & $\bs{1}$              & $1/3$        & $u^b_L$    \\
$V^{\dagger}_{12}$ & $\bs{3}^b$ & $\bs{2}^{\downarrow}$ & $\bs{1}$              & $1/3$        & $d^b_L$    \\
$V^{\dagger}_{13}$ & $\bs{1}$   & $\bs{1}$              & $\bs{2}^{\downarrow}$ & $-1$         & $e_R$      \\
$V^{\dagger}_{14}$ & $\bs{1}$   & $\bs{1}$              & $\bs{2}^{\uparrow}$   & $-1$         & $\nu_{eR}$ \\
$V^{\dagger}_{15}$ & $\bs{1}$   & $\bs{2}^{\uparrow}$   & $\bs{1}$              & $-1$         & $\nu_{eL}$ \\
$V^{\dagger}_{16}$ & $\bs{1}$   & $\bs{2}^{\downarrow}$ & $\bs{1}$              & $-1$         & $e_L$      \\
\hline
\end{tabular}
\caption{Transformation properties of each ideal under right actions of the group $G_{I}$. We define the colour multiplets $\bs{3}^r = (1,0,0)^T$, $\bs{3}^g = (0,1,0)^T$, $\bs{3}^b = (0,0,1)^T$ and the isospin doublets $\bs{2}^{\uparrow} = (1,0)^T$, $\bs{2}^{\downarrow} = (0,1)^T$. Although we chose to use the notation for the first family flavours, each ideal $V^{\dagger}_A$ must be multiplied by $\xi_{\al,g}$ from the left in order to fully specify the family and spinor indexes of the full multiplet. Since right multiplications only affect the flavour indexes, these results are replicated for each family and spinor indexes.}\label{Table1}
\end{table}
\begin{table}[h!]
\centering
\renewcommand{\arraystretch}{1.4}
\begin{tabular}{|c|c|c|c|}
\hline
\multicolumn{4}{|c|}{Representations of $\xi_{\alpha,g} V^{\dagger}_A$ under left action of $G_{E}$} \\
\hline
Multiplet & $Spin(1,3)$ & $SU(3)_F$ & $U(1)_F$ \\
\hline
$\xi_{a,1}V^{\dagger}_{A}$ & $(\frac{1}{2},0)$ & $\overline{\bs{3}}^1$ & $1/3$ \\
$\xi_{\tilde{a},1}V^{\dagger}_{A}$ & $(0,\frac{1}{2})$ & $\overline{\bs{3}}^1$ & $1/3$ \\
$\xi_{a,2}V^{\dagger}_{A}$ & $(\frac{1}{2},0)$ & $\overline{\bs{3}}^2$ & $1/3$ \\
$\xi_{\tilde{a},2}V^{\dagger}_{A}$ & $(0,\frac{1}{2})$ & $\overline{\bs{3}}^2$ & $1/3$ \\
$\xi_{a,3}V^{\dagger}_{A}$ & $(\frac{1}{2},0)$ & $\overline{\bs{3}}^3$ & $1/3$ \\
$\xi_{\tilde{a},3}V^{\dagger}_{A}$ & $(0,\frac{1}{2})$ & $\overline{\bs{3}}^3$ & $1/3$ \\
$\xi_{a,4}V^{\dagger}_{A}$ & $(\frac{1}{2},0)$ & $\bs{1}$              & $-1$ \\
$\xi_{\tilde{a},4}V^{\dagger}_{A}$ & $(0,\frac{1}{2})$ & $\bs{1}$              & $-1$ \\
\hline
\end{tabular}
\renewcommand{\arraystretch}{1}
\caption{Transformation properties of each field multiplet $\xi_{\al,g} V^{\dagger}_A$ under left actions of the group $G_{E}$. We define the index $a=1,2$ and the basis vectors of the anti-fundamental representation $\overline{\bs{3}}^1 = (1,0,0)$, $\overline{\bs{3}}^2 = (0,1,0)$, $\overline{\bs{3}}^3 = (0,0,1)$ to distinguish from the color indexes of $SU(3)$ as acting from the right. The spinor indexes are partitioned as $\al \equiv(a,\tilde{a})$, where $a=1,2$ and $\tilde{a}=3,4$.}\label{Table2}
\end{table}
For future convenience we also remark that, defining the Witt basis of a Dirac algebra $\ma{C}\ell_{1,3}$
\ea{
K_+ & = \frac{\Gamma^0+\Gamma^3}{2}\,, \label{K0}\\
K_0 & = \frac{\Gamma^1+i \Gamma^2}{2}\,, \label{Kp}
}
we obtain a set of relations between left and right-handed multiplets summarised in Table~\ref{Table3}.
\begin{table}[h]
\centering
\renewcommand{\arraystretch}{1.4}
\begin{tabular}{c|cccccccc}
 & $\nu_{e R}$ & $e_R$ & $u^{c}_{R}$ & $d^{c}_{R}$ \\
\hline
$\nu_{e L}$ & $-K^{\dagger}_{0}$ & $K_{+}$ \\
$e_L$       & $K^{\dagger}_{+}$ & $K_{0}$ \\
$u^{c}_{L}$ & & & $-K^{\dagger}_{0}$ & $K_{+}$ \\
$d^{c}_{L}$ & & & $K^{\dagger}_{+}$ & $K_{0}$ \\
\end{tabular} \label{weakStructureG}
\caption{Algebraic connections of left and right doublets via right multiplication with elements of 4-dimensional Dirac algebra elements. As an example, the entry $(e_L, \nu_{e R})$ corresponds to the identity $V^{\dagger}_{e_L} = V^{\dagger}_{\nu_{e R}} K^{\dagger}_{+}$. Non-existing entries imply that the corresponding ideals are not related by combinations of $\Gamma^{\mu}$ alone. The subscript $c$ denotes any colour.}\label{Table3}
\end{table}
All generators from the groups $G_I$ and $G_E$ commute with $\Gamma^5$, so the field multiplets (\ref{MultipletW}) are naturally divided by left and right action of $P_{L,R}$. However, it is easy to see that multiplets of the form $P_R\xi_{\al,g}V^{\dagger}_AP_L$ or $P_L\xi_{\al,g}V^{\dagger}_AP_R$ are absent from the kinetic Lagrangian (\ref{LkinaA}): since $P_{L} \xi_{\al,g} V^{\dagger}_{A} P_R = -\Gamma^{5}P_{L}\xi_{\al,g}V^{\dagger}_{A}P_R\Gamma^{5}$, we have that 
\ea{
& \textrm{Tr}\left[P_LV_{A}\xi^{\dagger}_{\al,g} P_L \Gamma^{0}\Gamma^{\mu} P_R \xi_{\al,g}V^{\dagger}_{A}P_L\right] \nonumber \\
& = - \textrm{Tr}\left[P_LV_{A}\xi^{\dagger}_{\al,g} P_L \Gamma^{0}\Gamma^{\mu} P_R \xi_{\al,g}V^{\dagger}_{A}P_L\right]
}
and so any multiplet with mismatched external and internal chiralities will be canceled in the Lagrangian. Consequently, the only remaining combination is
\ea{\label{AllParticlesL}
\Psi(x) = & \sum^{4}_{g=1}\sum^{16}_{A=1}\sum^{4}_{\al=1} P_R \xi_{\al,g}V^{\dagger}_A P_R \, \psi_{\al,g,A}(x) \nonumber \\
& \hspace{5mm} + P_L \xi_{\al,g}V^{\dagger}_{A} P_L \, \psi_{\al,g,A}(x)
}
instead of (\ref{MultipletW}), such that the most general rotation leaving the Lagrangian invariant is
\eq{\label{MultipletTrans}
\Psi'(x) = e^{X_E} \Psi(x') e^{X_I}\,,
}
where $x'^{\mu} = \Lambda^{\mu}{}_{\nu} x^{\nu}$ is the Lorentz transformation associated to the $Spin(1,3)$ rotation within $G_E$ and $X_E,X_I$ are the generators of the groups $G_E$ and $G_I$. Taking the latter remarks into account and gauging the symmetry groups $G_I$ and $G_E$, we find a suitably gauge invariant Lagrangian (\ref{LkinaA}) in the form of
\eq{
\ma{L}^{\textrm{Kin.}}(x) = \textrm{Tr}\bigg[\Psi^{\dagger}(x) (i\Gamma^{0}\Gamma^{\mu}) D_{\mu} \Psi(x)\bigg]
}
where we define the covariant derivative
\eq{
D_{\mu}\Psi = \frac{1}{2}(D^{I}_{\mu}\Psi+D^{E}_{\mu}\Psi) \label{Dmu} \\
}
where
\ea{
D^{I}_{\mu}\Psi = & \dd_{\mu}\Psi-2i\Psi(g_s G_{\mu}^{a} \Lambda_{a} +g_R W_{R,\mu}^{i} T^{R}_{i} \nonumber \\
& + g_L W_{L,\mu}^{i} T^{L}_{i} +g_{BL} B_{B-L,\mu} Q_{B-L}) \label{DmuInt} \\
D^{E}_{\mu}\Psi = & \dd_{\mu}\Psi-2i(\tilde{g}_s \tilde{G}_{\mu}^{a} \Lambda_{a} +\om^{\mu}{}_{\al\be} \Sigma^{\al\be} \nonumber \\
& +\tilde{g}_{BL} \tilde{B}_{B-L,\mu} Q_{B-L}) \Psi \label{DmuExt}
}
with appropriately introduced spin 1 gauge fields for each generator. As is standard practice, we choose not to introduce an individual gauge coupling for the Lorentz gauge fields $\om^{\mu}{}_{\al\be}$, whose quantization will not be addressed in this work. The kinetic terms for each remaining gauge boson $A^{a}_{\mu}(x)$ with structure constants defined by $[T^a,T^b]=f^{ab}{}_{c} T^c$ and gauge coupling $\tilde{g}$ are written as usual in the form
\ea{
\hspace{-3.5mm} \ma{L}^{\textrm{Kin.}}_{\textrm{spin 1}} & = -\frac{1}{4} \textrm{Tr}\left[\hat{\bs{A}}_{\mu\nu} \hat{\bs{A}}^{\mu\nu}\right], \\
\hspace{-3.5mm} \hat{\bs{A}}_{\mu\nu} & = \dd_{\mu} \hat{\bs{A}}_{\nu} - \dd_{\nu} \hat{\bs{A}}_{\mu} + \tilde{g} [\hat{\bs{A}}_{\mu},\hat{\bs{A}}_{\nu}], \\
\hspace{-3.5mm} \hat{\bs{A}}_{\mu} & = \sum_{a} A^{a}_{\mu} T_a\,.
}

\begin{widetext}

\begin{figure}[h!]
  \centering
  \includegraphics[width=0.8\textwidth]{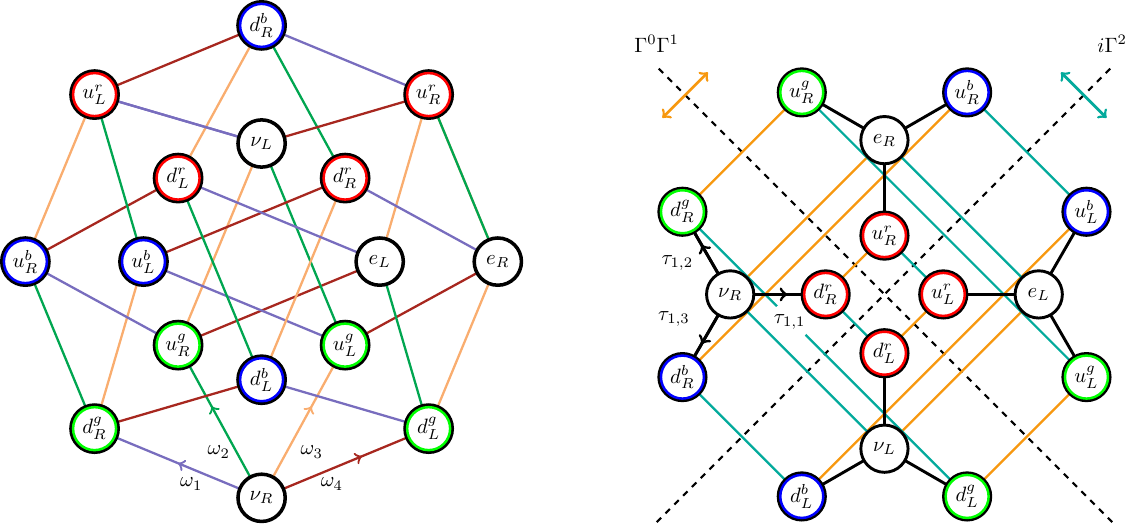}
  \caption{\textbf{Left:} right-multiplying the $\nu_{e R}$ ideal $V^{\dagger}_{14}$ by combinations of $\om_{\al}$ leads to all possible flavours (up to a sign). Upwards direction is associated to multiplication by $\om_{\al}$ and downwards by $\om_{\al}^{\dagger}$. \textbf{Right:} an alternative graph is obtained by right multiplication with combinations of spacetime vectors $\Gamma^0\Gamma^1$ and $i\Gamma^2$ (up to a sign) and Witt basis vectors $\tau_{\al,a}$ defined via $V^{\dagger}_{14} \tau_{\al,a} = \Omega^{\dagger}_{a} \om_{\al}$ (generating an algebra $\ma{C}\ell(6)$ for each $\al$). The multiplet structures of the weak and strong force become evident.}
  \label{VAfig}
\end{figure}

\end{widetext}

It is indispensable to emphasize that this theory is intrinsically connected to the algebraic structure of spacetime since the generators of $SU(2)_L \times SU(2)_R$ and its external counterpart $Spin(1,3)$ are functions of the observed 4-dimensional Clifford algebra, while those of $SU(3)\times U(1)_{B-L}$ are functions of the extra-dimensional directions and the 4-dimensional pseudoscalar. The latter statement can be straightforwardly verified by defining four additional basis vectors
\eq{
\Upsilon^{\bar{\mu}} = \tilde{\gamma}^{\bar{\mu}} \otimes \gamma^5\,, \quad \bar{\mu} = 1,\ldots,4\,,
}
where $\tilde{\gamma}^{\bar{\mu}}$ are the Dirac matrices normalized to $+1$, with which we can express the $SU(3)$ and $U(1)_{B-L}$ generators as
\ea{
Q_{B-L} & = \frac{1}{3}(-i \Upsilon^2\Upsilon^3 -i \Upsilon^1\Upsilon^4+\Upsilon^1\Upsilon^2\Upsilon^3\Upsilon^4)\,, \\
\Lambda^1 & = \frac{i}{2}(\Upsilon^{1}\Upsilon^{2}-\Upsilon^{3}\Upsilon^{4}) \,, \\
\Lambda^2 & = \frac{i}{2}(\Upsilon^{1}\Upsilon^{3}+\Upsilon^{2}\Upsilon^{4}) \,, \\
\Lambda^3 & = \frac{i}{2}(\Upsilon^{1}\Upsilon^{4}-\Upsilon^{2}\Upsilon^{3}) \,, \\
\Lambda^4 & = -\frac{1}{2}\Gamma^5(\Upsilon^{1}-i\Upsilon^{1}\Upsilon^{2}\Upsilon^{3}) \,, \\
\Lambda^5 & = \frac{1}{2}\Gamma^5(\Upsilon^{4}-i\Upsilon^{2}\Upsilon^{3}\Upsilon^{4}) \,, \\
\Lambda^6 & = \frac{1}{2}\Gamma^5(\Upsilon^{3}+i\Upsilon^{1}\Upsilon^{3}\Upsilon^{4}) \,, \\
\Lambda^7 & = \frac{1}{2}\Gamma^5(\Upsilon^{2}+i\Upsilon^{1}\Upsilon^{2}\Upsilon^{4}) \,, \\
\Lambda^8 & = \frac{1}{2\sqrt{3}}(i \Upsilon^2\Upsilon^3 +i \Upsilon^1\Upsilon^4+2\Upsilon^1\Upsilon^2\Upsilon^3\Upsilon^4)\,.
}
To close this section, we note that the structure of internal flavour space has a particular geometric form since one can move between $V^{\dagger}_A$ via right multiplication, leading to the graphs of Fig.~\ref{VAfig}.

\section{Spontaneous symmetry breaking of internal symmetries}
\label{Sec5}

With Tables~\ref{Table1}, \ref{Table2} and \ref{Table3} in hand, we can now break the symmetries responsible for the mass terms. Note, however, that while the fermionic sector is fully fixed, the scalar sector is entirely unconstrained, so the breaking chains and Higgs representations used are not unique. The main goal of this section is to show how scalar multiplets find a natural home in Clifford algebras, with corresponding spacetime interpretations.

Let us start with the breaking $SU(4)_C \to SU(3)_C \times U(1)_{B-L}$ in order to offer a dynamical picture complementary to that of the triality-induced breaking developed in the previous section. Following standard Pati-Salam theory and using the $SU(4)$ generators $T^k$ defined in Eq.~(\ref{SU4gens}), we introduce a Higgs $\Sigma = \Sigma^{k} T_{k} \otimes 1_{4\times 4} \sim (\bs{15},\bs{1},\bs{1})$ under $SU(4)_C \times SU(2)_L \times SU(2)_R$ (do not confuse with the $\Sigma^{\mu\nu}$ Lorentz generators (\ref{LorentzGens})) with gauge covariant derivative
\eq{
D_{\mu} \Sigma = \dd_{\mu} \Sigma - i g_4 [A^{k}_{\mu}T_{k}\otimes 1_{4\times 4} ,\Sigma]\,,
}
where
\eq{\label{SU4gaugeBosons}
A^{k}_{\mu}T_{k} =
\begin{pmatrix}
G_{\mu} + \frac{1}{2\sqrt{6}} B_{B-L,\mu} & X_{\mu} \\
X^{\dagger}_{\mu} & -\frac{3}{2\sqrt{6}} B_{B-L,\mu}
\end{pmatrix}
}
contains the eight gluons $G_{\mu}$, the $B-L$ gauge boson $B_{B-L,\mu}$ and the leptoquark gauge bosons $X_{\mu}$. The Lagrangian is of the form
\eq{
\ma{L}_{\Sigma} = \textrm{Tr}[(D_{\mu} \Sigma)^{\dagger}(D_{\mu} \Sigma)] - V(\Sigma)
}
with the most general renormalizable potential
\eq{\label{VSigma}
V(\Sigma) = \frac{\mu^2_{\Sigma}}{2} \textrm{Tr}[\Sigma^2]-\frac{a_{\Sigma}}{3} \textrm{Tr}[\Sigma^3] - \lambda_1 (\textrm{Tr}[\Sigma^2])^2 - \lambda_2 \textrm{Tr}[\Sigma^4]\,.
}
After $\Sigma$ develops the VEV
\eq{
\braket{\Sigma} = v_{\Sigma} T^{15} \otimes 1_{4\times 4}\,,
}
the unbroken generators are those associated to $G_{\mu}$ and $B_{B-L,\mu}$, corresponding to the generators (\ref{SU3gens}) and (\ref{Qgens}) of $SU(3)_C \times U(1)_{B-L}$. The advantage of instead inducing this $SU(4)_C$ breaking by fixing a triality is a considerable reduction in free parameters and particle content. Different triality choices simply correspond to different VEVs $\braket{\Sigma}$ related by a similarity transformation of the $SU(4)_C$ generators.

Moving on to the breaking $SU(2)_R\times U(1)_{B-L} \to U(1)_Y$, we can use the Higgs multiplet $\Delta_R \sim (\bs{1},\bs{1},\bs{3},+2)$ under $G_I$, responsible for the same breaking in the Pati-Salam theory \cite{Mohapatra:1980}, here represented by
\ea{
\Delta_R & = \frac{\Delta_{R}^{+}}{\sqrt{2}} \, V_{\nu_{eR}} V^{\dagger}_{\nu_{eR}} - \frac{\Delta_{R}^{+}}{\sqrt{2}} \, V_{e_R} V^{\dagger}_{e_R} \nonumber \\
& \hspace{5mm} + \Delta_{R}^{++} \, V_{\nu_{eR}} V^{\dagger}_{e_R} + \Delta_{R}^{0} \, V_{e_R} V^{\dagger}_{\nu_{eR}}  \,, \\
& \hspace{-5mm} = 
\begin{pmatrix}
0 & 0 & 0 & 0 \\
0 & 0 & 0 & 0 \\
0 & 0 & 0 & 0 \\
0 & 0 & 0 & 1
\end{pmatrix}
\otimes
\begin{pmatrix}
-\Delta_{R}^{+}/\sqrt{2} & \Delta^0_R & 0 & 0 \\
\Delta^{++}_R & \Delta_{R}^{+}/\sqrt{2} & 0 & 0 \\
0 & 0 & 0 & 0 \\
0 & 0 & 0 & 0
\end{pmatrix}\,,
}
with gauge covariant derivative
\eq{
D_{\mu} \Delta_R = \dd_{\mu} \Delta_R - i g_RW^{i}_{R,\mu}[T^{R}_{i}, \Delta_R]-2ig_{B-L} B_{B-L,\mu} \Delta_R\,.
}
Defining the operator
\eq{
i \tau_2 = \begin{pmatrix}
0 & 0 & 0 & 0 \\
0 & 0 & 0 & 0 \\
0 & 0 & 0 & 0 \\
0 & 0 & 0 & 1
\end{pmatrix}
\otimes
\begin{pmatrix}
0 & -1 & 0 & 0 \\
1& 0 & 0 & 0 \\
0 & 0 & 0 & 0 \\
0 & 0 & 0 & 0
\end{pmatrix}\,,
}
and the charge conjugate operator $C = i \Gamma^2\Gamma^0$ allows the introduction of the Majorana mass term
\eq{
\ma{L}_{\textrm{Maj.}} = \textrm{Tr}[f_R \Psi^{T}_{R,g} C \Psi_{R,g} (i\tau_2) \Delta_R]
}
where we define the chiral doublets of a single family
\ea{
\Psi_{L,g} = \sum_{\al} P_L\xi^{\dagger}_{\al,g}V_{g_L} g_L + P_L\xi^{\dagger}_{\al,g}V_{\nu_{gL}}\nu_{gL}\,, \label{LDoublet} \\
\Psi_{R,g} = \sum_{\al} P_R\xi^{\dagger}_{\al,g}V_{g_R} g_R + P_R\xi^{\dagger}_{\al,g}V_{\nu_{gR}}\nu_{gR}\,. \label{RDoublet}
}
Once the VEVs $\braket{\Delta^{+}_R} = \braket{\Delta^{++}_R} = 0$ and $\braket{\Delta^0_R} = v_R/\sqrt{2}$ are acquired, a Majorana mass $M_N = f_R v_R/\sqrt{2}$ appears and the gauge bosons $W^3_R$ and $B_{B-L}$ mix into the massless boson $B$ associated to the unbroken hypercharge operator
\eq{
Y = Q_{B-L} \otimes 1_{4\times 4} + 1_{4\times 4}\otimes (2T^{R}_3)\,.
}
The combination $i \tau_2 \braket{\Delta_R} = \ma{V}$ appearing in the right-doublet is exactly the quantity used to construct the QFT vacuum. So, just as fixing a triality breaks $SU(4)_C$ into $SU(3)_C \times U(1)_{B-L}$, fixing the vacuum $\ma{V}$ of the annihilation operators $\om^{\al}$ breaks $SU(2)_R\times U(1)_{B-L}$ into $U(1)_{Y}$.

Finally, we need to introduce the Higgs $\Phi$ responsible for the breaking $SU(2)_L \times U(1)_Y \to U(1)_Q$. We consider the standard choice of a bidoublet $\Phi_{PS} \sim (\bs{1},\bs{2},\bs{2})$ under $SU(4)_C \times SU(2)_L \times SU(2)_R$, commonly appearing in the literature as \cite{Branco:1999}
\eq{
\Phi_{PS} = (\Phi_{PS,1} \, \Phi_{PS,2}) = 
\begin{pmatrix}
\phi^{0 *}_1 & \phi^+_{2} \\
-\phi^-_{1} & \phi^0_2
\end{pmatrix}\,,
}
where $\Phi_{PS} \to U_L \Phi_{PS} U^{\dagger}_R$. Using Table~\ref{Table3}, it is straightforward to see that the map to a Clifford algebra representation is
\eq{\label{PhiPSCliff}
\Phi_{PS} \to \Phi = \Phi_1 + \Phi_2\,,
}
where 
\ea{
\Phi_1 & = P_R(-\phi^-_{1}K_{+}+\phi^{0 *}_1 K^{\dagger}_0)P_L\,, \\
\Phi_2 & = P_R(\phi^{+}_{2} K^{\dagger}_{+}  + \phi^0_2 K_0)P_L\,,
}
with gauge covariant derivative
\eq{
D_{\mu} \Phi = \dd_{\mu} \Phi +ig_L  \Phi W_{L,\mu}^{i} T^{L}_{i} - ig_R  W_{R,\mu}^{i} T^{R}_{i} \Phi\,,
}
where the actions of $U_L$ and $U_R$ are inverted due to internal transformations taking place as transformations from the right. After $\Delta_R$ acquires a VEV, $\Phi$ transforms as two electroweak doublets with opposite hypercharges under the SM group. Further fixing the VEV (ignoring CP-phase in the scalar sector for simplicity)
\eq{
\braket{\Phi_{PS}} =
\begin{pmatrix}
\kappa_1 & 0 \\
0 & \kappa_2
\end{pmatrix}
}
leads to the electric charge unbroken generator $Q = T^{L}_3 + \frac{Y}{2}$. Coupling unification is not achieved here but might happen within $\ma{C}\ell_{8,0}$. This would require rescaling $Y$ by $\sqrt{5/3}$ to match the normalization of $T^L_i$, giving the Weinberg angle $\sin^2\theta_W = 3/8$ at unification scale, as in standard GUTs \cite{Georgi:1974a,Georgi:1974b,Fritzsch:1975}. Defining $\tilde{\Phi} = \Gamma^2 \Gamma^0 \Phi^* \Gamma^2 \Gamma^0$ we can express the SM Higgs doublet in terms of $\Phi_1$ and $\Phi_2$ as the hypercharge $+1$ combination
\eq{
\Phi_{SM} = -\frac{\kappa_1}{\sqrt{2}v} \tilde{\Phi}_1 + \frac{\kappa_2}{\sqrt{2}v}\Phi_2
}
with the acquired VEV
\eq{
\braket{\Phi_{SM}} = \frac{v}{\sqrt{2}} P_R K_0 P_L
}
displaying a connection to spacetime null vectors in 4-dimensional spacetime since $K^2_0=0$. We will now focus on the family sector where new insights can be obtained.

\section{Symmetry breaking of external symmetries}
\label{Sec6}

As previously noted, the external real slice of (\ref{genXigroup}) selects $SU(4)_F \times Spin(1,3)$. While $Spin(1,3)$ is known to be an exact symmetry to great accuracy, $SU(4)_F$ must be broken to generate Yukawa matrices. One appealing approach is to introduce a Higgs multiplet $\bss{\varphi} \sim \bs{4} \otimes \overline{\bs{4}}$ under $SU(4)_F$ (a generalization of \cite{Koide:1990}), coupled to fermions within the families of a single flavour via
\eq{\label{U3FMass}
\ma{L}^{\textrm{Mass}}_{f} = \textrm{Tr}\left[\Psi_{L}^{\dagger}\Gamma^0\bss{\varphi}\bss{\varphi}^{\dagger}\Psi_{R}\Phi\right] + \textrm{h.c.}
}
where $\Phi$ is the bidoublet from Eq.~(\ref{PhiPSCliff}). Eq.~(\ref{U3FMass}) is treated as an effective, unrenormalizable term, unsurprising since higher dimensional terms should eventually emerge from the still-unavailable full 8-dimensional picture. Breaking $SU(4)_F$ via a specific triality choice in $\ma{C}\ell_{8,0}$ is appealing since it can simultaneously break the internal $SU(4)_C$ into $SU(3)_C \times U(1)_{B-L}$. If $SU(4)_F$ is instead broken spontaneously, a potential like (\ref{VSigma}) and a suitable VEV for $\bss{\varphi}$ are needed instead. Either way, due to the lack of an observed fourth family so far, we expect the result to be the family symmetry $SU(3)_F \times U(1)_F$, under which $\bss{\varphi}$ decomposes as
\eq{
\bss{\varphi} =
\begin{pmatrix}
\bss{\varphi}_T  & \bss{\varphi}_X \\
\bss{\varphi}^{\dagger}_X & \varphi_9
\end{pmatrix} \otimes 1_{4\times 4}\,, \quad 
\bss{\varphi}_T = \bss{\varphi}_* + \frac{\varphi_0}{\sqrt{3}} 1_{3\times 3}
}
such that, under $SU(3)_F \times U(1)_F$, $\bss{\varphi}_* = \varphi_k \lambda^k \sim (\bs{8},0)$, $\varphi_0 \sim (\bs{1},0)$, $\varphi_9 \sim (\bs{1},0)$, $\bss{\varphi}_X \sim (\bs{3},\frac{4}{3})$ and $\bss{\varphi}^{\dagger}_X \sim (\overline{\bs{3}},-\frac{4}{3})$. The multiplets $\bss{\varphi}_X$ and $\bss{\varphi}^{\dagger}_X$ are either eaten by gauge bosons that become very massive, or forbidden outright by the algebraic constraint of fixing a triality, so mixing between the fourth family and the other three should be extremely small or exactly zero. The symmetry $SU(3)_F \times U(1)_{F}$ must itself be broken by a Higgs potential, which we consider to be (generalizing \cite{Koide:1990})
\ea{
V(\bss{\varphi}) & = \mu^2_9 |\varphi_9|^2 - \lambda_9 |\varphi_9|^4 + \mu^2 \textrm{Tr}[\bss{\varphi}_T\bss{\varphi}^{\dagger}_T] - \lambda (\textrm{Tr}[\bss{\varphi}_T\bss{\varphi}^{\dagger}_T])^2 \nonumber \\
& \hspace{5mm} - \nu |\varphi_0|^2|\varphi_9|^2 - \lambda'  |\varphi_0|^2 \textrm{Tr}[\bss{\varphi}_* \bss{\varphi}^{\dagger}_*]\,. \label{U3HiggsPotential}
}
This is not the most general renormalizable potential (as noted in \cite{Koide:1990}), which could be obtained by breaking $V(\Sigma)$ in Eq.~(\ref{VSigma}), but it suffices to showcase some features directly impacting the Yukawa matrix structure. The conditions $\dd V(\bss{\varphi}) / \dd \varphi_k = 0$, $\dd V(\bss{\varphi}) / \dd \varphi_0 = 0$ and $\dd V(\bss{\varphi}) / \dd \varphi_9 = 0$, which must be satisfied for the potential minimum, result in the VEVs
\eq{\label{varphiVEVs}
\braket{\varphi_0} = v_{\varphi} \,, \quad \braket{\varphi_9} = v_9\,,
}
where
\eq{
v^2_{\varphi} = \frac{\la_9 \la' \mu^2 +  \mu^2_9 \la \nu}{\la_9 \la' (4\la+\la') + \la \nu^2}\,, \quad v^2_{9} = \frac{\mu^2_9}{2\lambda_9}-\frac{\nu}{2\lambda_9} v^2_{\varphi}  
}
with the VEVs of $\varphi_0$ and $\varphi_9$ taken as real, a redefinition allowed by $V(\bss{\varphi})$. Most interestingly, the VEVs $\braket{\bss{\varphi}_T}$ also satisfy
\eq{\label{NewKoide}
\frac{\textrm{Tr}[\braket{\bss{\varphi}_T}\braket{\bss{\varphi}^{ \dagger}_T}]}{\textrm{Tr}[\braket{\bss{\varphi}_T}]\textrm{Tr}[\braket{\bss{\varphi}^{\dagger}_T}]} = \frac{2}{3} - \Delta\,, \quad \Delta = \frac{\nu}{3\la'} \frac{v^2_9}{v^2_{\varphi}}
}
which, as shown in the next section, directly encodes a generalized Koide formula corrected by $\Delta$. The kinetic term $\textrm{Tr}[D_{\mu}\bss{\varphi}^{\dagger}D^{\mu}\bss{\varphi}]$ gives mass to all $U(3)_F$ gauge bosons, which must be large to suppress tree-level flavour-changing neutral currents (FCNC). Although we will not pursue the phenomenological details of further constraining $\braket{\bss{\varphi}_T}$, we can still conclude that Yukawa matrices will have the general form
\eq{\label{VEVmatrix}
\ma{Y} = \braket{\bss{\varphi}}\braket{\bss{\varphi}^{\dagger}} =
\begin{pmatrix}
Y_f  & 0 \\
0 & y_f^{(4)}
\end{pmatrix} \otimes 1_{4\times 4}
}
where $Y_f$ is a general 3-dimensional hermitian matrix and $y_f^{(4)}$ a real number.

A sequential fourth family is well known to be in tension with experiment, but since our framework predicts one unavoidably, the issue deserves discussion. Four main obstacles arise \cite{Navas:2025} from the Yukawa coupling $y_f^{(4)}$: 1) Z boson decay constraints; 2) cosmological constraints; 3) electroweak oblique correction constraints; and 4) constraints from Higgs production via gluon fusion. The first two are avoided if the fourth-family neutrinos are too heavy to be produced in Z decays, meaning they also annihilate away well before big bang nucleosynthesis;  there is room for this here via the Majorana mass terms. The fourth problem arises because the gluon-fusion Higgs production amplitude with a heavy fourth family is proportional to $\ma{A}(gg\to H) \propto \kappa_t + \kappa_{t'} + \kappa_{b'}$ with primes used for fourth family particles and $\kappa_f = g_{hff}/g^{SM}_{hff}$, with $g_{hff}$ and $g^{SM}_{hff}$ the generic and SM Higgs couplings, respectively. If all $\kappa_f = +1$, the predicted amplitude is roughly 3 times the observed one, which is ruled out experimentally. An elegant fix is to extend the scalar sector so $\kappa_{t'}$ and $\kappa_{b'}$ have opposite signs. Certain type-II Two-Higgs-Doublet Models (2HDM) \cite{Carmi:2012,Chiang:2013,Ferreira:2014,Fontes:2014,Das:2018} exhibit this "wrong-sign limit" while also solving the electroweak oblique correction constraints. These models, however, have been mostly ruled out by non-perturbative corrections from their inevitable extra heavy Higgs sectors \cite{Kang:2018}. Nevertheless, while the Type-II 2HDM route is now strongly constrained, this does not exclude all realizations of a sequential fourth family as alternative scalar sectors may yet work. Thus, while the fourth-family hypothesis lacks appeal as a standalone SM extension, it is by no means ruled out.

\section{A parameterization of hermitian mass matrices}
\label{Sec7}

The model in Eq.~(\ref{U3FMass}) gives hermitian Yukawa matrices across all flavours, leading to mass matrices $i M \in \mathfrak{u}(3)$ for the first three families in the non-mixing fourth-family limit. Interestingly, since hermitian matrices have a real spectrum, the positivity of $M$ can be fully parameterized via (c.f. Appendix \ref{App3})
\eq{\label{MW}
M = W^2, \quad W^{\dagger} = W\,.
}
The structure of Eq.~(\ref{MW}) results in the useful parameterization
\ea{
\sqrt{m_k} & = \frac{2E}{\sqrt{3}}\left|\cos\chi + \sqrt{2}\sin \chi \cos\left(\de + \frac{2\pi k}{3}\right)\right|\,, \label{sqrtm} \\
E & = \frac{\sqrt{m_1+m_2+m_3}}{2}\,, \label{Emass}\\
\cos \chi & = \frac{\sum_k\textrm{sign}(\varepsilon_k)\sqrt{m_k}}{\sqrt{3}\sqrt{m_1+m_2+m_3}}\,, \label{coschi}\\
\cos(3\de) & = \frac{1}{4\sqrt{2}\sin^3\chi}\bigg(24\sqrt{3} \sqrt{\frac{m_1m_2m_3}{(m_1+m_2+m_3)^3}} \nonumber \\
& \hspace{10mm} -3\cos\chi -5\cos(3\chi) \bigg)\,, \label{newdelta}
}
where $k=0,1,2$, $(\chi,\de)$ are angles, and $\textrm{sign}(\varepsilon_k)$ is the sign of $W$'s $k$th eigenvalue $\varepsilon_k$. Eq.~(\ref{coschi}) resembles the Koide relation \cite{Koide:1982}, but with signed square-root masses and the Koide constant $3/2$ parameterized as $3\cos^2 \chi$. Comparing (\ref{VEVmatrix}) and (\ref{MW}) gives $\braket{\bss{\varphi}_T}\braket{\bss{\varphi}^{\dagger}_T} \propto M = W^2$ which, when inserted in Eq.~(\ref{NewKoide}), results in
\eq{\label{genKoide}
\frac{(m_1+m_2+m_3)}{\left(\sum_k\textrm{sign}(\varepsilon_k)\sqrt{m_k}\right)^2} = \frac{2}{3} - \Delta
}
or, using the parameterization (\ref{coschi}), $\cos^2 \chi = 1/(2-3\Delta)$. As $\Delta \to 0$, $\chi = \pi/4$, reducing the right side of Eq.~(\ref{genKoide}) to the standard Koide formula. Inserting the latest charged lepton masses \cite{Navas:2025} into Eq.~(\ref{newdelta}) (at fixed $\chi=\pi/4$) gives
\eq{
\delta = 0.222221(2)\,,
}
very close to the simple value $\delta=\frac{2}{9}$, which fixes all signs in Eq.~(\ref{genKoide}) as positive, recovering the full standard Koide formula. The angles $(\chi,\de)=(\pi/4,2/9)$ also appear in mass models using circulant matrices and cascade breaking \cite{Brannen:2006,Goffinet:2008}. For neutrinos, \cite{Brannen:2006,Mohapatra:2006} found masses consistent with $(\chi,\de)=(\pi/4,2/9-\pi/12)$, requiring a postulated sign change in Koide's square roots that arises naturally from Eq.~(\ref{coschi}). For charged leptons, the right side of Eq.~(\ref{genKoide}) being so close to $2/3$ implies $\nu/\la' = 3\Delta(v^2_{\varphi}/v^2_9)$ must be exceedingly small, since the heavy fourth family already forces $(v^2_{\varphi}/v^2_9) \ll 1$. For quarks, current mass precision leaves $(\chi,\de)$ too uncertain to determine.

We finish by noting that, for fixed $\chi$, $E$ and $\delta$ can be expressed as functions of a single parameter $\tau_{\chi} \in \mathbb{C}$, since the characteristic equation for $W$ is solved by elliptic functions (c.f. Appendix \ref{App3}). In particular, $E^2$ (the mass scale) is the modular function
\eq{\label{Etau}
E^2(\tau_{\chi}) = \frac{g_2(\tau_{\chi})}{4\sin^2\chi}
}
with $g_2(\tau)$ the second Weierstrass invariant, while $\delta$ relates to the modular lambda function $\lambda(\tau)$ via
\eq{\label{tandelta}
\tan \delta(\tau_{\chi}) = \frac{\sqrt{3}}{1-\frac{2}{\lambda(\tau_{\chi})}}\,.
}
Equations (\ref{Etau})-(\ref{tandelta}) suggest mass matrices are fundamentally connected to modular symmetries (see \cite{Kobayashi:2024,Novichkov:2024} for recent reviews of this hypothesis).

\section{Conclusions}
\label{Sec8}

We showed that the SM (modulus the scalar sector), plus additional beyond-SM phenomena, follows from the free Dirac Lagrangian in 8-dimensional spacetime. Motivated by the expression of fixed-momentum fermionic second-quantized operators as Clifford algebra elements, we generalized the free Dirac Lagrangian from $\ma{C}\ell_{1,3}$ to $\ma{C}\ell_{8,0}$. This gave four copies of 4-dimensional Dirac spinors, tied to the adjoint and three fundamental representations of Spin(8), predicting four particle families and naturally representing field multiplets as ideals of $\ma{C}\ell_{8,0}$. These multiplets carry mathematical redundancies under $G = SL(4,\mathbb{C}) \times Spin(4)_{\mathbb{C}}$, with the invariance of the Dirac Lagrangian under $G$ selecting two real slices, $G_E = SU(4)_F \times Spin(1,3)$ and $G_I = SU(4)_C\times SU(2)_L \times SU(2)_R$, when field multiplet actions are restricted to left (external) or right (internal) multiplication, respectively. Fixing a triality in $\ma{C}\ell_{8,0}$ was further found to simultaneously break $SU(4)_C \to SU(3)_C \times U(1)_{B-L}$ and $SU(4)_F \to SU(3)_F \times U(1)_F$ with both abelian parts generated by the same $B-L$ generator. It is therefore tempting to speculate that the existence of the leptonic sector may be linked to a hypothetical non-mixing sequential fourth family, whose lightest neutrino could provide a viable dark matter candidate. Spontaneous breaking of internal and external symmetries was analyzed from a Clifford algebra viewpoint, giving spacetime algebraic interpretations for each VEV. The Higgs used to break $U(3)_F$ gives hermitian Yukawa matrices satisfying a generalized Koide formula, with the first three family masses parameterized by two angles $(\chi,\de)$ and a positive real $E$, with the choice $(\pi/4,2/9)$ found to reproduce the charged lepton mass ratios with striking accuracy. Once $\chi$ is fixed, $(E,\de)$ is parametrized by a single complex number, giving each family flavour's mass scale a modular nature.

Though not set in stone, the extension to $\ma{C}\ell_{8,0}$ opens up the possibility of at least four additional dimensions. Fortunately, the details of the higher dimensional metric were not necessary to obtain the SM. It is unknown if the extra dimensions are spacelike, timelike or a mixture of both. Since the SM Lagrangian is expressed in terms of 4-dimensional dynamics, some dimensional reduction mechanism must be operating, though which one is unclear. This should affect quantum fields as functions of 8-dimensional spacetime coordinates, allowing new higher dimensional Lagrangian terms such as (\ref{U3FMass}), favoring higher dimensions as a source of new physics and making a fully 8-dimensional Dirac equation the natural next step. The family mixing problem in particular clearly connects to the extra four dimensions, given the shared structure between the mass matrices and the strong symmetry generators—both functions of the extra-dimensional spacetime basis vectors—suggesting the same mechanism integrating the extra dimensions could be potentially linked to the mass matrices.

The harshest constraints of this theory undoubtedly fall on the fermionic sector, allowing only a predicted fourth family and right-handed neutrinos for all families, severely restricting possible SM extensions. Constraints on the bosonic sector remain less understood, however, with Higgs field multiplets fitting comfortably in $\ma{C}\ell_{8,0}$ with no evident restrictions. Coupling unification is not yet achieved, since the couplings have no a priori relation. A fully dynamical, not just algebraic, picture in higher dimensional spacetime may change this.

\section*{Acknowledgements}

The author thanks the support from Funda\c{c}\~{a}o para a Ci\^{e}ncia e a Tecnologia through projects CEECIND/02474/2018 and 2024.04456.CERN. The author acknowledges everyone who gave feedback regarding the state-of-the-art, in particular Niels Gresnigt, after an earlier preprint version. The author also thanks Filipe Joaquim for useful discussions, and Nikolay Ketsaris for independently checking the results and for discussions regarding matrix representations. The author thanks the hospitality of the Maryland Center for Fundamental Physics where part of this research was developed as a visiting researcher, supported by the grant FCT-Mobility/1312232346/2024-25.


\clearpage
\pagebreak
\newpage

\widetext

\begin{center}
\textbf{\large Appendix for ``Spacetime Grand Unified Theory''}
\end{center}

\appendix


\section{Clifford algebra representations of second quantised fermionic operators}
\label{App1}

In this section we present the explicit matrix construction used for $e^k$, $\omega^\al$ and the second quantised fermionic creation/annihilation operators. We begin with the basis
\ea{\label{eAbasis}
& \sum^{8}_{k=1} c_k e^k = 
\begin{pmatrix}
0 & 1 \\
0 & 0
\end{pmatrix}\otimes
\left(
\begin{array}{cccccccc}
 c^{++}_{72} & 0 & c^{+-}_{54} & c^{-+}_{63} & c^{-+}_{81} & 0 & 0 & 0 \\
 c^{--}_{81} & 0 & 0 & 0 & c^{-+}_{72} & 0 & c^{+-}_{54} & c^{-+}_{63} \\
 c^{--}_{54} & c^{-+}_{63} & c^{+-}_{72} & 0 & 0 & 0 & c^{-+}_{81} & 0 \\
 c^{++}_{63} & c^{-+}_{54} & 0 & c^{+-}_{72} & 0 & 0 & 0 & c^{-+}_{81} \\
 0 & c^{++}_{72} & c^{++}_{63} & c^{++}_{54} & 0 & c^{+-}_{81} & 0 & 0 \\
 0 & c^{--}_{81} & 0 & 0 & 0 & c^{+-}_{72} & c^{++}_{63} & c^{++}_{54} \\
 0 & 0 & c^{--}_{81} & 0 & c^{--}_{54} & c^{+-}_{63} & c^{--}_{72} & 0 \\
 0 & 0 & 0 & c^{--}_{81} & c^{++}_{63} & c^{+-}_{54} & 0 & c^{--}_{72} \\
\end{array}
\right) \nonumber \\
& +
\begin{pmatrix}
0 & 0 \\
1 & 0
\end{pmatrix}\otimes
\left(
\begin{array}{cccccccc}
 c^{+-}_{72} & c^{-+}_{81} & c^{-+}_{54} & c^{+-}_{63} & 0 & 0 & 0 & 0 \\
 0 & 0 & c^{--}_{63} & c^{--}_{54} & c^{+-}_{72} & c^{-+}_{81} & 0 & 0 \\
 c^{++}_{54} & 0 & c^{++}_{72} & 0 & c^{+-}_{63} & 0 & c^{-+}_{81} & 0 \\
 c^{--}_{63} & 0 & 0 & c^{++}_{72} & c^{+-}_{54} & 0 & 0 & c^{-+}_{81} \\
 c^{--}_{81} & c^{--}_{72} & 0 & 0 & 0 & 0 & c^{-+}_{54} & c^{+-}_{63} \\
 0 & 0 & 0 & 0 & c^{++}_{81} & c^{++}_{72} & c^{++}_{63} & c^{++}_{54} \\
 0 & c^{++}_{54} & c^{--}_{81} & 0 & 0 & c^{+-}_{63} & c^{-+}_{72} & 0 \\
 0 & c^{--}_{63} & 0 & c^{--}_{81} & 0 & c^{+-}_{54} & 0 & c^{-+}_{72} \\
\end{array}
\right)
}
where we defined $c^{s_1  s_2}_{i,j} \equiv s_1 c_i+s_2 c_j$. For the construction of the Witt basis in Eq.~(\ref{WittCl8}), the lowest dimensional representation are $16 \times 16$ matrices belonging to $\ma{C}\ell_{8,0}$ with multiple possible metric signatures. Since the algebra will be complexified, we will take $\om^{\al} \in \ma{C}\ell_{8,0}$ without loss of generality, which can be constructed from (\ref{explicitWitt}) and (\ref{eAbasis}) with the matrices $\om^{\al\dagger}$ obtained via conjugate transposition.

Finally, we show how second quantized creation/annihilation operators can be expressed in terms of Clifford algebras. Start by discretizing space as a cubic lattice in three spatial dimensions with length $L$ and lattice spacing $a$. In momentum space, this corresponds to discretizing momentum into $N=(L/a)^3$ modes, labeled $\bs{p}_I$ with $I=(i_1,i_2,i_3)$ $(i_n \in \{1,2,\ldots,N\})$, leading to the discretized anticommutation relations
\ea{
& \{a^{s}_{\bs{p}_I}, a^{r \dagger}_{\bs{p}_J}\} = \{b^{s}_{\bs{p}_I}, b^{r \dagger}_{\bs{p}_J}\} = L^3 \de^{sr}\de^{i_1 j_1}\de^{i_2 j_2}\de^{i_3 j_3}\,, \label{latticeACR1} \\
& \{a^{s}_{\bs{p}_I}, a^{r}_{\bs{k}_J}\} = \{a^{s \dagger}_{\bs{p}_I}, a^{r \dagger}_{\bs{k}_J}\} = \{b^{s}_{\bs{p}_I}, b^{r}_{\bs{k}_J}\} = \{b^{s \dagger}_{\bs{p}_I}, b^{r \dagger}_{\bs{k}_J}\} = 0 \label{latticeACR2}
}
which define a Witt basis. We recall a theorem due to Cartan \cite{Cartan:1908}, commonly referred to as Bott periodicity, which states that $\ma{C}\ell_{l,n+8} \simeq \ma{C}\ell_{l,n}\otimes \ma{C}\ell_{0,8}$ or $\ma{C}\ell_{l+8,n} \simeq \ma{C}\ell_{l,n}\otimes \ma{C}\ell_{8,0}$. This property implies that the set
\eq{
\{\Lambda^i\} = \{ \om^{s_1}\otimes I_e \otimes \cdots \otimes I_e, 1_{16\times 16} \otimes \om^{s_2} \otimes I_e \otimes \cdots \otimes I_e, \ldots, 1_{16\times 16} \otimes \cdots \otimes 1_{16\times 16} \otimes \om^{s_N}\}
}
defines a Witt basis of $\ma{C}\ell_{8N,0}$ $(N\in \mathbb{N}^{+})$, where $s_k = 1,\ldots,4$ and each $\Lambda^{i}$ is a $16^{N}$-dimensional matrix. Since
\eq{
\{\Lambda^i,\Lambda^{j \dagger}\} = \de^{ij}\,, \quad \{\Lambda^i,\Lambda^j\} = \{\Lambda^{i \dagger},\Lambda^{j \dagger}\} = 0 \quad (i,j=1, 2,\ldots,N) 
}
one finds that the correspondence
\ea{
a^{1}_{\bs{p}_I} & \equiv L^{3/2}(\om^{1} \otimes \Lambda^{i_1} \otimes \Lambda^{i_2} \otimes \Lambda^{i_3})\,, \label{a1} \\
a^{2}_{\bs{p}_I} & \equiv L^{3/2}(\om^{2} \otimes \Lambda^{i_1} \otimes \Lambda^{i_2} \otimes \Lambda^{i_3})\,, \label{a2} \\
b^{1}_{\bs{p}_I} & \equiv L^{3/2}(\om^{3} \otimes \Lambda^{i_1} \otimes \Lambda^{i_2} \otimes \Lambda^{i_3})\,, \label{b1} \\
b^{2}_{\bs{p}_I} & \equiv L^{3/2}(\om^{4} \otimes \Lambda^{i_1} \otimes \Lambda^{i_2} \otimes \Lambda^{i_3})\,, \label{b2}
}
obeys Eqs.~(\ref{latticeACR1})-(\ref{latticeACR2}). In the limit $N \to \infty$, one recovers $a^{s}_{\bs{p}_I} \to a^{s}_{\bs{p}}$ and $b^{s}_{\bs{p}_I} \to b^{s}_{\bs{p}}$. Consequently, $a^{s}_{\bs{p}},b^{s}_{\bs{p}} \in \ma{C}\ell_{8,0}\otimes \lim_{N \to \infty} (\ma{C}\ell_{8N,0})^{\otimes 3}$ or, due to Bott periodicity,
\eq{
a^{s}_{\bs{p}},b^{s}_{\bs{p}} \in \ma{C}\ell_{8,0}\otimes \lim_{N \to \infty} (\ma{C}\ell_{8,0})^{\otimes 3N}
}
and, since quantum fields are linear combinations of creation and annihilation operators, we also have that $\psi_{\al}(x) \in \ma{C}\ell_{8,0}\otimes \lim_{N \to \infty} (\ma{C}\ell_{8,0})^{\otimes 3N}$. Since the quantized momentum modes are also generated by $\om^{\al}$, we find that the full QFT vacuum in the continuous limit can be written as the matrix
\eq{\label{QFTvac}
\ket{0} = V^{\dagger}_A \otimes \lim_{N \to \infty} V_A^{\dagger \otimes 3N}
}
for any fixed $A$. Although real algebras could be considered at this stage, they must eventually be complexified ultimately because the Dirac equation involves both an imaginary unit and the Dirac matrices. This is why we consider the complexified algebra of $\ma{C}\ell_{8,0}$ from the beginning.

One may rightfully argue that the Bott periodicity of any complexified algebra is 2 instead of 8, and thus Eqs.~(\ref{a1})-(\ref{b2}) could equally well be achieved with tensor products of $\ma{C}\ell_{2,0}$, for example. However, this overlooks the argument that $\ma{C}\ell_{8,0}$ (modulus metric signatures) is the simplest algebra containing precisely the right number of creation/annihilation operators for a single momentum mode. Although this does not constitute a mathematical proof of the uniqueness of our choice, it strengthens the physical motivation to consider $\ma{C}\ell_{8,0}$ as the algebra to focus on.

Finally, we note that in \cite{Furey:2021} it is observed that Bott periodicity might be related to the construction of multiparticle states and possibly Fock spaces. This is then used as motivation to study $\ma{C}\ell_{0,8}$ as previous works using the same algebra \cite{Gillard:2019a,Gillard:2019b,Gourlay:2024,Gresnigt:2026a} (modulus metric signatures), i.e. by searching for structures whose invariant symmetries might lead to the SM. While sharing the focus on a $\ma{C}\ell_{0,8}$ algebra (or Clifford algebras in general) as the setting to develop a theory of fundamental forces, this latter approach is distinct from the one developed in this work. In particular, the use of $\ma{C}\ell_{8,0}$ here is physically motivated by second quantization, as shown by the novel proofs in this section (which happen to use the Bott periodicity theorem). The references most closely aligned to this work are \cite{Zenczykowski:2015}, where (first) quantization is connected to $\ma{C}\ell_{6,0}$ and \cite{Cho:1995,Pavsic:2017}, which considered explicit representations of fermionic quantum fields, living in the algebras $\ma{C}\ell_{1,3}\otimes \ma{C}\ell_{\infty}$ and $\ma{C}\ell_{8,0}\otimes \ma{C}\ell_{\infty}$, respectively. The proposal that the SM symmetries and particle representations might be embedded in $\ma{C}\ell_{8,0}$ or $\ma{C}\ell_{0,8}$ goes at least as far back as 2010, in the concluding remarks of \cite{Pavsic:2010}.


\section{Four-dimensional Dirac spinor representations in $\ma{C}\ell_{8,0}$}
\label{App2}

We present a derivation of Eqs.~(\ref{g0gmuTrFinal})-(\ref{g0TrFinal}), as well as an introduction to triality in $\ma{C}\ell_{8,0}$. The Spin($n$) group has three fundamental representations: two spinorial ones, denoted by $S_{\pm}$ and a vectorial one, denoted by $V$. For $n=8$, all these representations are 8-dimensional, so any two elements of different representations uniquely define a third one in the remaining representation - a phenomenon called triality.

Consider an element $v = v^{k} e_{k}$ of the vector space of $\ma{C}\ell_{8,0}$, denoted by $V_{\ma{C}\ell}$. Let $s^{(0)} \in \ma{C}\ell_{8,0}$ be an even product of normalized vectors of the form $s^{(0)} = v_{(1)} v_{(2)} \dots v_{(2r)}$ where $v_{(k)} \in V_{\ma{C}\ell}$ are vectors such that $v_{(k)}v_{(k)}=1$. The objects $s^{(0)}$ are commonly referred to as spinors. Then $s^{(0)}$ will rotate the vector $v$ into ${g^{(0)}(v) = s^{(0)} v (s^{(0)})^{-1} \in V_{\ma{C}\ell}}$ where $g^{(0)} \in V$. Now take a vector $n \in V_{\ma{C}\ell}$ and an element $w \in \ma{C}\ell_{8,0}$, made of linear combinations of $e_i e_j e_k e_l$, such that
\eq{
n^2 = 1\,, \quad w^2 = 7+6w\,, \quad [w,n] = 0 \label{weq}\,, \\
}
known as calibration $w$ and its neutral axis $n$. Denoting $\ma{C}\ell^+_{8,0}$ as the subspace of $\ma{C}\ell_{8,0}$ made of even products, we can generate two independent spinor spaces $\ma{S}_{\pm}$ via \cite{Lounesto:2001}
\eq{
\ma{S}_{\pm} = \ma{C}\ell^+_{8,0} \left(\frac{1+w}{8}\right) \left(\frac{1\pm I_e}{2}\right)\,.
}
Elements $g^{(\pm)}$ of Spin(8) in the spinor representations $S_{\pm}$ will act as
\eq{\label{spinorRotation}
g^{(\pm)}(\ma{S}_{\pm}) = s^{(\pm)}\ma{C}\ell^+_{8,0} \left(\frac{1+w}{8}\right) \left(\frac{1\pm I_e}{2}\right)\,,
}
where $s^{(\pm)}$ are spinors. Since $s^{(\pm)} \in \ma{C}\ell^+_{8,0}$, then $s^{(\pm)}\ma{C}\ell^+_{8,0} \in \ma{C}\ell^+_{8,0}$ and so $g^{(\pm)}(\ma{S}_{\pm}) \in \ma{S}_{\pm}$ represent rotations in the spaces $\ma{S}_{\pm}$, which are connected to rotations in $V_{\ma{C}\ell}$ by \cite{Lounesto:2001}
\eq{\label{rot3spaces}
\si^{(0)} v (\si^{(0)})^{-1} = \braket{\si^{(\pm)}v(1+w)(1 \pm I_e)}_1
}
where $v \in V_{\ma{C}\ell}$, $\braket{x}_1$ is the vector part of $x \in \ma{C}\ell_{8,0}$ and $\si^{(0)}, \si^{(\pm)}$ are spinors, related by
\ea{
\si^{(-)} & = n\textbf{trial}(n \si^{(+)}n)n \,, \\
\si^{(0)} & = \textbf{trial}(\textbf{trial}(n \si^{(+)}n))
}
where $\textbf{trial}$ is the triality operation, defined for spinors $u$ as
\ea{
\textbf{trial}(u) & = \textbf{trial}_1(u)\textbf{trial}_2(u) \,, \label{trial} \\
\textbf{trial}_1(u) & = \left(\frac{1+I_e}{2}\right)\{\braket{u(1+w)(1-I_e)}_{0,6},n\}\frac{n}{2} +\left(\frac{1-I_e}{2}\right)\,, \\
\textbf{trial}_2(u) & = (w-3)\{u(1+I_e),n\}\frac{n}{2}(w-3)^{-1}\,,
}
where $\braket{x}_{0,6}\equiv\braket{x}_{0}+\braket{x}_{6}$ represents the summation of the scalar part of $x \in \ma{C}\ell_{8,0}$ and the component made of linear combinations of $e_i e_j e_k e_l e_m e_n$. The triality operation obeys the property ${\textbf{trial}(\textbf{trial}(\textbf{trial}(u)))=u}$ for any spinor $u$. Using \cite{Lounesto:2001}
\eq{
\braket{x}_1 = \braket{x e_k}_0 e_k = \frac{1}{16} \textrm{Tr}[x e_k]e_k
}
and $\textrm{Tr}[e_i e_j] = 16 \, \de_{ij}$, one can obtain an alternative form of Eq.~(\ref{rot3spaces}) by considering $v = e_k$, multiplying by $e_i$ from the right and taking the trace, resulting in
\ea{
\textrm{Tr}\left[\frac{\si^{(0)} e_k (\si^{(0)})^{-1} e_i}{16}\right] = \textrm{Tr}\left[e_i\si^{(\pm)}e_k\left(\frac{1+w}{8}\right)\left(\frac{1\pm I_e}{2}\right)\right]\,. \label{rot3spacesAB}
}
To explore the consequences of Eq.~(\ref{rot3spacesAB}) for the matrix components of a Dirac algebra element $X_{\al\be}$, we use Eqs.~(\ref{X8D}) and (\ref{XalbeV1}) to obtain 
\eq{\label{XPplus}
X_{\al\be} = \textrm{Tr}[\om_{\al} P_{(+)} X_{(\om)} P_{(+)} \om^{\dagger}_{\be}]\,,
}
and define the object $\Omega_{+}$ via
\eq{\label{OPlusDef}
\om_{\al} P_{(+)} = \Omega^{\dagger}_{+} \om_{\al}
}
which, due to Eq.~(\ref{XPplus}), relates to $X_{\al\be}$ through $X_{\al\be} = \textrm{Tr}[\om_{\al} X_{(\om)} \om^{\dagger}_{\be}\Omega_{+}\Omega^{\dagger}_{+}]$. Comparing with Eq.~(\ref{XalbeV1}), we find that $\Omega_{+}\Omega^{\dagger}_{+}$ should be proportional to an idempotent, in order to span a spinor space like $\ma{V}$ does. In other words, we should have
\eq{\label{OPlusIdem}
\left(\Omega_{+}\Omega^{\dagger}_{+}\right)^2 = C \, \Omega_{+}\Omega^{\dagger}_{+}\,, \quad C \in \mathbb{R}.
}
Using the representation (\ref{eAbasis}), Eqs.~(\ref{OPlusDef}) and (\ref{OPlusIdem}) uniquely lead to
\ea{\label{OPlusGen} 
\Omega_{+} = (\rho e^{-i \phi} V_{9} + V_{14})V^{\dagger}_{14}\,, \quad C = 1+\rho^2\,, \quad \rho>0 \in \mathbb{R}\,.
}
The ideal $\Omega_{+}$ also represents an algebraic vacuum, since $\om_{\al} \Omega^{\dagger}_{+} = 0$. Noting that
\eq{
X_{\al\be} = (\om_{\al})^i (\om^{\dagger}_{\be})^k \textrm{Tr}[e_i X_{(\om)} e_k \Omega_{+}\Omega^{\dagger}_{+}]
}
one can make contact with Eq.~(\ref{rot3spacesAB}) by requiring that $\Omega_{+}\Omega^{\dagger}_{+}$ is proportional to one of the idempotents $\left(\frac{1+w}{8}\right)\left(\frac{1\pm I_e}{2}\right)$. We can use the representation (\ref{eAbasis}) to find that only $\left(\frac{1+w}{8}\right)\left(\frac{1+I_e}{2}\right)$ has a compatible form, with the additional constraint $\rho = 1$, leading to
\eq{\label{OPlusFinal}
\Omega_{+} = (e^{-i \phi} V_{9} + V_{14})V^{\dagger}_{14}\,,
}
which implies that
\eq{\label{OPlusFormulas}
\left(\Omega_{+}\Omega^{\dagger}_{+}\right)^2 = 2 \, \Omega_{+}\Omega^{\dagger}_{+}\,, \quad \Omega^{\dagger}_{+}\Omega_{+} = 2 \ma{V}\,.
}
Since $\Omega_{+}\Omega^{\dagger}_{+}$ is proportional to $\left(\frac{1+w}{8}\right)\left(\frac{1+I_e}{2}\right)$, (\ref{OPlusFormulas}) leads to
\eq{\label{OOPlusfinal}
\Omega_{+}\Omega^{\dagger}_{+} = 2\left(\frac{1+w}{8}\right)\left(\frac{1+I_e}{2}\right) \,,
}
independently of $w$ and $n$. Eq.(\ref{OPlusFinal}) can also be expressed as
\eq{\label{wThetaPlus}
\Omega_{+} = \left(\frac{1+w}{8}\right)\left(\frac{1+ I_e}{2}\right)\Theta_{+}
}
where $\Theta_{+}$ is an object which, due to (\ref{OPlusFormulas}), must satisfy
\eq{\label{O32Formula}
\left(1+w\right)\left(1+I_e\right) \Theta_{+} \Theta^{\dagger}_{+} \left(1+w\right)\left(1+I_e\right) = 32 \left(1+w\right)\left(1+I_e\right)\,.
}
We can now introduce a quantity $\Omega_{-}$ satisfying (\ref{OPlusFormulas}) but associated instead to the idempotent $\left(\frac{1+w}{8}\right)\left(\frac{1-I_e}{2}\right)$, leading to an analogous version of (\ref{OOPlusfinal})
\eq{\label{OOMinusfinal}
\Omega_{-}\Omega^{\dagger}_{-} = 2\left(\frac{1+w}{8}\right)\left(\frac{1-I_e}{2}\right) \,.
}
Using (\ref{weq}) one immediately finds via (\ref{OOPlusfinal}) that
\eq{\label{OOMP1}
n\Omega_{+}\Omega^{\dagger}_{+}n = 2\left(\frac{1+w}{8}\right)\left(\frac{1-I_e}{2}\right)=\Omega_{-}\Omega^{\dagger}_{-}
}
so $\Omega_{-}\Omega^{\dagger}_{-}$ becomes uniquely defined once $\Omega_{+}$ and $n$ are fixed. Adding (\ref{OOPlusfinal}) and (\ref{OOMP1}) and solving for $w$, we also find 
\eq{\label{wGenFinal}
w = 4 \, \Omega_{+}\Omega^{\dagger}_{+} + 4 \, n\Omega_{+}\Omega^{\dagger}_{+}n -1\,.
}
Finally, similarly to $\Omega_{+}$, one can express $\Omega_{-}$ as
\eq{\label{wThetaMinus}
\Omega_{-} = \left(\frac{1+w}{8}\right)\left(\frac{1-I_e}{2}\right)\Theta_{-}
}
where $\Theta_{-}$ must satisfy Eq.~(\ref{O32Formula}) with $I_e \to -I_e$. With no loss of generality, in this work we have chosen $\phi = 0$, $n = e_8$, $\Theta_{+} = 2n\om^{\dagger}_{1}$ and $\Theta_{-} = 2\om^{\dagger}_{1}$ which, through Eq.~(\ref{wGenFinal}), leads to the calibration
\ea{
w = e_{1234} + e_{1467} - e_{1357} + e_{1256} - e_{3456}  - e_{2457} - e_{2367}\,, \label{calibration}
}
where we define $e_{ijkl} \equiv e_i e_j e_k e_l$. We now focus on the class of Dirac matrices such that $N_{\al}{}^{\nu} N_{\nu\be} = \de_{\al \be}$. For this class of matrices one finds, through Eq.~(\ref{X8D}), that
\eq{\label{Nabspinor}
P_{(+)} N_{(\om)} P_{(+)} = P_{(+)}\left(-i e^{i \frac{\pi}{2} N_{(B)}}\right)P_{(+)} \equiv -i P_{(+)}u^{(+)}P_{(+)}
}
where $N_{(B)} \equiv \braket{N_{(\om)}}_2$ is the component made of linear combinations of $e_k e_i$, thus making $u^{(+)}=e^{i \frac{\pi}{2} N_{(\om)}}$ a spinor by definition \cite{Lounesto:2001}. Using (\ref{rot3spacesAB}), we find that
\ea{
N_{\al\be} & = -2i \, \textrm{Tr}\left[\frac{u^{(0)} \om^{\dagger}_{\be} (u^{(0)})^{-1} \om_{\al}}{16}\right] \label{NabV} \\
& = -2i \, \textrm{Tr}\left[u^{(\pm)} \om^{\dagger}_{\be} \left(\frac{1+w}{8}\right)\left(\frac{1\pm I_e}{2}\right)\om_{\al}\right] \label{NabSpm}
}
where ${u^{(-)} = n\textbf{trial}(n u^{(+)} n)}n$ and ${u^{(0)} = \textbf{trial}(\textbf{trial}(n u^{(+)} n))}$. Using Eqs.~(\ref{wThetaPlus}) and (\ref{wThetaMinus}), we get
\ea{\label{NabOmPM}
N_{\al\be} = -i \, \textrm{Tr}\left[\Omega^{\dagger}_{\pm}\om_{\al}u^{(\pm)} \om^{\dagger}_{\be} \Omega_{\pm}\right]\,,
}
which begs the question: can a quantity $\Omega_{0}$ be defined such that (\ref{NabV}) can be expressed as (\ref{NabOmPM})? Focusing on the explicit cases at hand, namely $\g^0 \g^{\mu}$ and $\g^0$, it follows that
\ea{
(\g^0)_{(B)}          & = \frac{i}{2}(e_4e_7 +e_3e_8+e_2e_5+e_1e_6) \,, \\
(1_{4\times 4})_{(B)} & = -\frac{i}{2}(e_4e_5 + e_3e_6+e_2e_7+e_1e_8) \,, \\
(\g^0\g^{1})_{(B)}    & = \frac{i}{2}(e_4e_6 +e_3e_5-e_2e_8-e_1e_7) \,, \\
(\g^0\g^{2})_{(B)}    & = \frac{i}{2}(e_5e_6 -e_7e_8-e_3e_4+e_1e_2) \,, \\
(\g^0\g^{3})_{(B)}    & = \frac{i}{2}(e_3e_6-e_4e_5 +e_2e_7-e_1e_8) \,.
}
This implies that Eq.~(\ref{NabV}) just needs to be applied to the spinors
\ea{
u^{\mu}_{(0)} & \equiv \textbf{trial}\left[\textbf{trial}\left[n \exp\left(i \frac{\pi}{2}(\g^0 \g^{\mu})_{(B)}\right) n\right]\right]\,, \label{umu0} \\
\si_{(0)}  & \equiv \textbf{trial}\left[\textbf{trial}\left[n \exp\left(i \frac{\pi}{2}(\g^0)_{(B)}\right) n\right]\right]\,. \label{usi0}
}
It thus suffices to find an ideal $\Omega_{0}$ which, in addition to satisfying the properties in (\ref{OPlusFormulas}), also obeys $(u^{\mu}_{(0)})^{-1} \Omega_{(0)} = \Omega_{(0)}$ and $(\si_{(0)})^{-1} \Omega_{(0)} = \Omega_{(0)}$. The unique solution is
\eq{\label{OmZero}
\Omega_{0} = e^{i \varphi}(-\om_1\om_2+\om_3\om_4)\om^{\dagger}_{1}\om^{\dagger}_{2}\om^{\dagger}_{3}\om^{\dagger}_{4}\,,
}
which, due to the explicit use of the triality operation in Eqs.~(\ref{umu0})-(\ref{usi0}), requires a calibration and neutral axis. We choose $\varphi=0$ in this work, without loss of generality. Using Eqs.~(\ref{NabV}), (\ref{NabOmPM}) and (\ref{OmZero}), one thus arrives at the representations
\ea{
(\g^0 \g^{\mu})_{\al\be} & = -i \, \textrm{Tr}\left[\Omega^{\dagger}_{a} \om_{\al} u^{\mu}_{(a)} \om^{\dagger}_{\be} \Omega_{a}\right]\,, \label{g0gmuTr} \\
(\g^0)_{\al\be} & = -i \, \textrm{Tr}\left[\Omega^{\dagger}_{a} \om_{\al} \si_{(a)} \om^{\dagger}_{\be} \Omega_{a}\right]\,, \label{g0Tr}
}
where $a\in \{+,-,0\}$. There is a final fourth representation which can be fixed now that we have access to the ideals $\Omega_{a}$, associated to the projector
\ea{\label{PperpCond}
P_{(4)} & = 1_{16 \times 16} - \sum_{a,\al} \om^{\dagger}_{\al}\Omega_a\Omega^{\dagger}_a \om_{\al} \\
& = \textrm{diag}(0,0,0,0,0,0,0,0,0,0,0,0,1,1,1,1) \,,
}
which cannot be generated by the ideals $\om^{\dagger}_{\al}\Omega_a$. The projector $P_{(4)}$ must be associated to the remaing representation of Spin(8), the adjoint, represented by Lie algebra elements $X = X^{ij} [e_i,e_j]$, with associated Dirac ideals living in
\eq{\label{AdjRep}
X_{(4)} = P_{(4)} X P_{(4)} = 0_{12\times 12} \oplus
\begin{pmatrix}
x_{1,1} & x_{1,2} & x_{1,3} & x_{1,4} \\
x_{2,1} & x_{2,2} & x_{2,3} & x_{2,4} \\
x_{3,1} & x_{3,2} & x_{3,3} & x_{3,4} \\
x_{4,1} & x_{4,2} & x_{4,3} & x_{4,4}
\end{pmatrix}\,.
}
The simplest way to obtain similar expressions to Eqs.~(\ref{g0gmuTr})-(\ref{g0Tr}) for the adjoint representation is to directly map 4-dimensional Dirac matrices into (\ref{AdjRep}), leading to
\ea{
\si_{(4)} & = 0_{12\times 12} \oplus \g^0 = P_{(4)} \left(\frac{i}{2} [e_4,e_7]\right) P_{(4)}\,, \\
u^{0}_{(4)} & = 0_{12\times 12} \oplus 1_{4\times 4} = P_{(4)} \left(\frac{i}{2} [e_8,e_1]\right) P_{(4)}\,, \\
u^{1}_{(4)} & = 0_{12\times 12} \oplus \g^0\g^1 = P_{(4)} \left(\frac{i}{2} [e_4,e_6]\right) P_{(4)}\,, \\
u^{2}_{(4)} & = 0_{12\times 12} \oplus \g^0\g^2 = P_{(4)} \left(\frac{i}{2} [e_4,e_3]\right) P_{(4)}\,, \\
u^{3}_{(4)} & = 0_{12\times 12} \oplus \g^0\g^3 = P_{(4)} \left(\frac{i}{2} [e_5,e_4]\right) P_{(4)}\,.
}
A similar set of equations (\ref{g0gmuTr})-(\ref{g0Tr}) can now be written as
\ea{
(\g^0 \g^{\mu})_{\al\be} & = -i \, \textrm{Tr}\left[\kappa^{\dagger}_{\al} u^{\mu}_{(4)} \kappa_{\be}\right]\,, \label{g0gmuTrPerp} \\
(\g^0)_{\al\be} & = -i \, \textrm{Tr}\left[\kappa^{\dagger}_{\al} \si_{(4)} \kappa_{\be}\right]\,, \label{g0TrPerp}
}
with (up to a global sign change)
\eq{
\hspace{-0.8mm} \kappa_{1} = V_{13} \equiv \xi_{1,4}\,, \,\,\, \kappa_{2} = V_{14} \equiv \xi_{2,4}\,, \,\,\, \kappa_{3} = V_{15} \equiv \xi_{3,4}\,, \,\,\, \kappa_{4} = V_{16} \equiv \xi_{4,4}\,.
}
Crucially, $\xi_{\al,4}=\kappa_{\al}$ cannot be written in the form $\om^{\dagger}_{\al} \Omega_{4}$ for any $\Omega_{4} \in \ma{C}\ell_{8,0}$. This finishes the proof of Eqs.~(\ref{xi1})-(\ref{xi4}). The index $(a)$ in $u^{\mu}_{(a)}$ and $\si_{(a)}$ can now be removed by using instead the set of four matrices $\Gamma^{\mu} = 1_{4 \times 4} \otimes \gamma^{\mu} \in C\ell_{8,0}$ in the traces of Eqs.~(\ref{g0gmuTr})-(\ref{g0Tr}) and (\ref{g0gmuTrPerp})-(\ref{g0TrPerp}), simplifying Eqs.~(\ref{g0gmuTr})-(\ref{g0Tr}) into
\ea{
(\g^0 \g^{\mu})_{\al\be} & = \textrm{Tr}\left[\xi^{\dagger}_{\al,g} \Gamma^{0}\Gamma^{\mu} \xi_{\be,g}\right]\,, \nonumber \\
(\g^0)_{\al\be} & = \textrm{Tr}\left[\xi^{\dagger}_{\al,g} \Gamma^{0} \xi_{\be,g} \right]\,, \nonumber
}
using the definition of $\xi_{\al,g}$ in Eqs.~(\ref{xi1})-(\ref{xi4}). Using $\xi_{\al,g} \xi^{\dagger}_{\be,g} = \xi_{\al,g} \ma{V} \xi^{\dagger}_{\be,g}$ and $\ma{V} = V^{\dagger}_A V_A$ in the above equations results in Eqs.~(\ref{g0gmuTrFinal})-(\ref{g0TrFinal}), thereby finishing their proof.

\section{Proof of the family mass matrices parameterization}
\label{App3}

In this section we will provide a concise proof for Eqs.~(\ref{sqrtm})-(\ref{newdelta}). From Sec.~\ref{Sec6}, we obtain a mass matrix for the first three families parameterized as a 3-dimensional hermitian matrix with real positive semi-definite eigenvalues, non-degenerate in general. If $W$ is a general 3-dimensional hermitian matrix, then
\eq{\label{MH2}
M = W^2
}
will satisfy all the latter requirements, since $W$ has a real eigenvalue spectrum. We will use the following parameterization
\eq{\label{Hdef}
W = q 1_{3\times 3} + H
}
where $q\in \mathbb{R}$, $H = \sum^{8}_{k=1} c_k \la_{k}$ $(c_k \in \mathbb{R})$ and $\la_k$ are the Gell-Mann matrices. The eigenvalues of $H$ are given by \cite{Curtright:2015}
\eq{\label{xik}
\varepsilon'_k = \frac{2 |\bs{c}|}{\sqrt{3}} \cos\left( \delta + \frac{2\pi k}{3} \right)\,, \quad k=0,1,2
}
with $|\bs{c}|^2 = \sum^8_{k=1} c^2_k$ and where $\delta$ is an angle uniquely defined by the invariant $\det(H)$ through the parametric equation
\eq{
\cos(3\delta) = \frac{3\sqrt{3}}{2|\bs{c}|^3} \det(H)\,.
}
Defining the variables $E$ and $\chi$ as
\ea{
E \sin \chi & = \frac{|\bs{c}|}{\sqrt{2}}\,, \label{echi1} \\
E \cos \chi & = \frac{q \sqrt{3}}{2}\,, \label{echi2}
}
and inserting in Eq.~(\ref{Hdef}), leads to the eigenvalues $\varepsilon_k$ of $W$ as
\eq{\label{epsilonk}
\varepsilon_k = \frac{2E}{\sqrt{3}}\left(\cos\chi + \sqrt{2} \sin\chi \cos\left( \delta + \frac{2\pi k}{3} \right)\right)\,, \quad k=0,1,2\,.
}
Equation~(\ref{MH2}) implies that each mass eigenvalue $m_k$ is given by $m_k = \varepsilon^2_k$, such that $|\varepsilon_k| = \sqrt{m_k}$, thereby proving Eq.~(\ref{sqrtm}). Since $\textrm{Tr}[M] = m_1+m_2+m_3$, one can use Eq.~(\ref{Hdef}) and Eq.~(\ref{echi2}) to recover Eq.~(\ref{Emass}). Similarly, the trace of Eq.~(\ref{Hdef}) with Eqs.~(\ref{echi2}) and (\ref{Emass}) leads to Eq.~(\ref{coschi}). Calculating $\det(M)$, using Eqs.~(\ref{MH2}) and (\ref{epsilonk}) and simultaneously noting that $\det(M)= m_1m_2m_3$, leads to Eq.~(\ref{newdelta}), thereby allowing one to calculate $\delta$ using $\chi$ and the masses $m_k$ without having to specify the eight degrees of freedom $c_k$. This finishes the proof of Eqs.~(\ref{sqrtm})-(\ref{newdelta}).

We now proceed to the proof of equations Eqs.~(\ref{Etau})-(\ref{tandelta}). The characteristic polynomial for $H$ is given by
\eq{\label{ellipticF}
4\varepsilon'^3 - g_2 \varepsilon' - g_3 = 0 
}
with $g_2 = 2 \textrm{Tr}\left[H^2\right]$, $g_3 = 4 \det\left(H\right)$ and eigenvalues (\ref{xik}). The solutions to equation (\ref{ellipticF}) are elliptic functions. The functions $g_2$ and $g_3$ are called Weirstrass invariants and are known to be parameterizable by a single complex coefficient $\tau$, exhibiting the properties of modular functions of weight 4 and 6, respectively \cite{Apostol:1990}. Using $\textrm{Tr}\left[H^2\right] = 2 |\bs{c}|^2$ in conjunction with Eqs.~(\ref{echi1}) and $g_2$ (for fixed $\chi$), we obtain Eq.~(\ref{Etau}). Finally, the definition of the modular lambda function in terms of the roots $\varepsilon'_k$ leads to
\eq{
\lambda(\tau) = \frac{\varepsilon'_2 - \varepsilon'_1}{\varepsilon'_0 - \varepsilon'_1} = \frac{2}{\frac{\sqrt{3}}{\tan \delta} +1}
}
which can be inverted to deduce Eq.~(\ref{tandelta}).

\end{document}